\documentclass[aps,prx,reprint, amsmath, amssymb,superscriptaddress]{revtex4-1}

\usepackage{bm}
\usepackage{graphicx}
\usepackage{mathrsfs}
\usepackage{amsmath}
\usepackage{amssymb,mathtools}
\usepackage{color}
\usepackage{ulem}
\usepackage{amsfonts}
\usepackage{times,txfonts}
\usepackage{nicefrac}
\usepackage[colorlinks=true,linkcolor=blue,urlcolor=blue,citecolor=blue,pdfusetitle]{hyperref}

\newcommand{\kB}{k_{\mbox{\tiny B}}}

\DeclareMathOperator{\csch}{csch}
\begin{document}

\title{Efficient asymmetric collisional Brownian particle engines}
\author{C. E. Fern\'andez Noa}
\affiliation{Instituto de F\'isica da Universidade de S\~ao Paulo,  05508-090 S\~ao Paulo, SP, Brazil}
\author{Angel L. L. Stable}
\affiliation{Instituto de F\'isica da Universidade de S\~ao Paulo,  05508-090 S\~ao Paulo, SP, Brazil}
\author{William G. C. Oropesa}
\affiliation{Instituto de F\'isica da Universidade de S\~ao Paulo,  05508-090 S\~ao Paulo, SP, Brazil}
\author{Alexandre Rosas}
\affiliation{Departamento de F\'isica, CCEN, Universidade Federal da Para\'iba, Caixa Postal 5008, 58059-900, Jo\~ao Pessoa, Brazil}
\author{C. E. Fiore}
\affiliation{Instituto de F\'isica da Universidade de S\~ao Paulo,  05508-090 S\~ao Paulo, SP, Brazil}
\date{\today}

\begin{abstract}
  The construction of efficient thermal engines operating  at finite times  constitutes a fundamental and timely topic
  in nonequilibrium thermodynamics. We introduce a strategy for optimizing the performance  of  Brownian engines, based on a collisional approach for unequal interaction times  between the system and  thermal reservoirs.
  General  (and exact) expressions for thermodynamic properties and their optimized values are obtained, irrespective of the driving forces, asymmetry, the temperatures of reservoirs and protocol to be maximized. Distinct routes for the  engine optimization, including
    maximizations of output power and efficiency  with respect to the asymmetry, force and both of them are investigated. For the isothermal work-to-work converter and/or small difference
  of temperature between reservoirs, they are solely  expressed   in terms
  of Onsager coefficients.
  Although the symmetric engine can operate very inefficiently depending on the control parameters,   the usage of distinct contact times between the system and each reservoir not only can enhance the machine performance
  (signed by  an  optimal tuning  ensuring the largest gain)
  but also enlarges substantially the machine regime operation.
  The present approach
  can pave the way for the construction of efficient Brownian engines operating
  at finite times.
\end{abstract}

\maketitle

\section{Introduction}

A long-standing dilemma in Thermodynamics and related areas concerns the issue of
mitigating the impact of thermal noise/wasted heat
in order to improve the machine performance.
This constitutes a high  relevant problem, not only for theoretical purposes but
also for the construction of  experimental setups \cite{callen,prigo,groot}. Giving that 
the machine performance is commonly   dependent on
particular chemical compositions and operation conditions, notably for small-scale engines, the role of fluctuations being crucial for such engines, distinct  approaches have been proposed
and investigated in the realm of  stochastic and quantum thermodynamics \cite{seifert2012stochastic,van2005thermodynamic}.
A second fundamental point concerns that, even if all sources of dissipation could be mitigated, the performance of any thermal machine would still be limited by Carnot efficiency, which requires the occurrence of infinitely slow quasi-static processes and consequently the engine operates at null power. In contrast, realistic systems operate at finite time and power. Such  conundrum
(control/mitigation of dissipation and engine optimization) has contributed for the discovery of several approaches based on the maximization of power output instead of the efficiency~\cite{verley2014unlikely, schmiedl2007efficiency, seifert2012stochastic, esposito2009universality, cleuren2015universality, van2005thermodynamic, esposito2010quantum, seifert2011efficiency, izumida2012efficiency, golubeva2012efficiency, holubec2014exactly, bauer2016optimal, karel2016, tu2008efficiency, ciliberto2017experiments,bonanca2019,rutten2009reaching}.

Thermal machines based on Brownian particles have been successfully studied not only for theoretical purposes \cite{verley2014unlikely,bauer2016optimal,schmiedl2007efficiency,proesmans2017underdamped}
but also for the building of reliable experimental setups  \cite{martinez2016brownian,proesmans16b,krishnamurthy2016micrometre,blickle2012realization,quinto2014microscopic,kumar2018nanoscale}. They are also remarkable for depicting 
the limitations of classical thermodynamics and disclose the scales
in which thermal fluctuations  become relevant. In several   situations, thermal machines  involve isothermal transformations  \cite{blickle2012realization,martinez2016brownian,proesmans16b}. Such class of processes are fundamental  in thermodynamic since they are minimally dissipative. However,  isothermal transformations  are slow, demanding  sufficient large number of stages for achieving the desired final state. For this reason,
distinct protocols, such as increasing the
coupling between system and the thermal bath,  have been
undertaken for speeding it  up and simultaneously controlling
the increase of dissipation \cite{PhysRevLett.105.150603,PhysRevLett.98.108301,PhysRevX.10.031015,PhysRevA.103.032211,PhysRevLett.124.110606}.

Here we introduce a strategy for  optimizing the performance of
irreversible Brownian machines operating at isothermal parts via the control
of interaction time between the system and the environment. 
 Our approach is based on 
a Brownian particle 
sequentially placed in contact with distinct
thermal baths and  subject to external forces \cite{noa2020thermodynamics}
for unequal  times.
Such description, also  referred as collisional,   has been
successfully employed in different contexts, such as
systems that interact  only
with a small fraction of the environment and those
presenting distinct drivings over each member of system \cite{benn1,maru,saga,parrondo}.
Depending on the parameters of the model (period, driving
and difference of temperatures),
the symmetric version can operate very inefficiently.
Our aim is to show that the machine performance improves substantially by tuning properly the interaction time between particle and each reservoir. Besides the increase of the power and/or efficiency, the asymmetry in the contact time also enlarges the regime of operation of the machine substantially. Contrasting
with previous works \cite{PhysRevLett.98.108301,PhysRevX.10.031015,PhysRevA.103.032211,PhysRevLett.124.110606},  the optimization is solely obtained
via the control of interaction time 
and no external parameters are considered.
We derive  general relations
for distinct kinds of maximization, including
the maximization of the efficiency and power  with respect to
the force,  the asymmetry and both of them. For the isothermal work-to-work converter and/or small difference
  of temperature between reservoirs, they are solely  expressed   in terms
  of Onsager coefficients.
  The present approach
  can pave the way for the construction of efficient Brownian engines
  operating at finite times.

This paper is organized as follows: In Sec. II we present the thermodynamic
 of Brownian particles subject to asymmetric time
switching. In Sec. III, the efficiency is analyzed for two cases:
the work-to-work converter  processes and distinct temperature reservoirs.  Optimization
protocols are presented and exemplified for distinct drivings. Finally, conclusions are drawn in Sec. IV and explicit calculations of Onsager coefficients and linear regimes are present in Appendixes.
\section{Thermodynamics of asymmetric interaction times}
We consider a Brownian particle with  mass $m$ sequentially and cyclically placed in contact with  different thermal reservoirs, each at a temperature $T_i$ for time interval $\tau_i$. Here $i=1,\ldots,N$ label the reservoirs and also the order of contact between the reservoirs and the particle. While in contact with the $i$-th reservoir, the velocity $v_i(t)$ of the  particle evolves in time according to the Langevin equation
\begin{equation}
    \frac{d v_i}{dt} =  - \gamma_i v_i +f_i(t)+\zeta_{i}(t),
  \label{two_baths_mov1}
\end{equation}
where
$\gamma_i,f_i(t)$ and $\zeta_i(t)$  denote, respectively, the viscous constants, external forces and stochastic forces (interaction between particle and the $i$-th reservoir), all divided by the mass of the particle. Stochastic forces are assumed to satisfy the white noise properties:
\begin{equation}
 \langle \zeta_{i}(t)\rangle=0,
  \label{two_baths_ruido1}
\end{equation}
and,
\begin{equation}
 \langle\zeta_{i}(t)\zeta_{i^{\prime}}(t^{\prime}) \rangle= 2 \gamma_i T_i \delta_{ii^{\prime}} \delta (t - t^{\prime}).
  \label{two_baths_ruido2}
\end{equation}

The system evolves to a nonequilibrium steady state regime ({\it NESS})
characterized by a non-vanishing production of entropy.
The time evolution of the velocity probability distribution at time $t$, $P_i(v,t)$, is described by the Fokker-Planck equation \cite{mariobook,tome2010,tome2015}
\begin{equation}
\frac{\partial P_i}{\partial t} = - \frac{\partial J_i}{\partial v} -f_i(t)\frac{\partial P_i}{\partial v}, 
\label{64}
\end{equation}
where $J_i$ is the probability current 
\begin{equation}
  \quad J_i = - \gamma_i v P_i - \frac{\gamma_i k_{B} T_i}{m}\frac{\partial P_i}{\partial v}.
  \label{645}
  \end{equation}
As can be verified by direct substitution, the {\it NESS} is characterized by a Gaussian  probability distribution $P_i(v,t)$:
\begin{equation}
P_i(v,t)=\frac{1}{\sqrt{2\pi b_i(t)}}e^{-\frac{(v-\langle v_i \rangle)^{2}}{2b_i(t)}},
\end{equation}
for which the mean $\langle v_i \rangle(t)$
and the variance $b_i(t)\equiv\langle v_i^2 \rangle(t) -\langle v_i \rangle^2(t)$ are time-dependent and obey the following equations of motion
\begin{equation}
  \frac{d}{dt}\langle v_i\rangle(t) =-\gamma_i\langle v_i\rangle(t) +f_i(t),
  \label{v1}
\end{equation}
and
\begin{equation}
      \frac{d}{dt} b_i(t)  = -  2 \gamma_i   b_i(t)  + \Gamma_i,
\label{v2}
\end{equation}
where  $\Gamma_i=2\gamma_i\kB T_i/m$. Obviously, the continuity of the probability distribution must be assured, and we will use it to calculate $b_i(t)$ and $\langle v_i \rangle (t)$ in the following sections.

In order to derive explicit expressions for macroscopic
quantities, we start from the definitions
of the average energy $U_i=  m \langle v_i^2\rangle/2$ and entropy $S_i(t) = - \kB \langle \ln [P_i(v,t)] \rangle$,
respectively. In both cases, the time variation   can be straightforwardly obtained from the Fokker-Planck equation and applying vanishing boundary conditions for both $P_{i}(v,t)$ and $J_i(v,t)$ in the infinity speed limit~\cite{mariobook}.
The former is related with the   average power dissipated $\dot{W}_i$ and the
heat dissipation during the same period $\dot{Q}_i$ through  the first law of thermodynamics relation:
\begin{equation}
  \frac{dU_i}{dt}=- [{\dot W_i(t)}+{\dot Q_i(t)}],
  \label{rele}
\end{equation}
where $\dot{W}_i(t)$ and $\dot{Q}_i(t)$ are given by the following expressions:
\begin{equation}
\dot{W}_i(t) = -m \langle v_i\rangle(t) f_i(t),
\label{112}
\end{equation}
and
\begin{equation}
\dot{Q}_i(t) = m\gamma_i\left(\langle v_{i}^{2}\rangle(t) -  \frac{\Gamma_i}{2\gamma_i}\right).
\label{112b}
\end{equation}
Similarly, the rate of variation of the entropy can be written as~\cite{tome2010,tome2015}:  
\begin{equation}
\frac{dS_i}{dt}=\Pi_i(t)-\Phi_i(t),
\end{equation}
where $\Pi_i(t)$ and $\Phi_i(t)$ denote the entropy production rate and the flux of entropy, respectively, which expressions are given by,
     \begin{equation}
  \Pi_i(t)=\frac{2\kB}{\Gamma_i} \int \frac{J_i^2}{P_i}dv,
     \end{equation}
     and
\begin{equation}
  \Phi_i(t)=- \frac{2 \gamma_i k_B}{\Gamma_i}\int v J_i dv =\frac{2\gamma_i\kB \dot Q_i(t)}{m\Gamma_i} = \frac{\dot{Q}_i(t)}{T_i}.
  \label{eqf}
\end{equation}
Both expression are valid during the contact of the Brownian particle with the $i$-th reservoir.  

As stated before, the present collisional approach for Brownian machines can be considered for an
arbitrary set of reservoirs and external forces, which generic solutions $ \langle v_i\rangle(t)$'s
and $b_i(t)$'s in the nonequilibrium steady state regime are 
\begin{equation}
  \langle v_i\rangle(t)=e^{-\gamma_i (t-\tilde{\tau}_{i-1})} a_{i}+ e^{-\gamma_i t} F_i(t),
  \label{eqvv}
\end{equation}
  and
\begin{equation}
  b_i(t)=A_i e^{-2\gamma_i (t-\tilde{\tau}_{i-1})}+\frac{\Gamma_i}{2\gamma_i},
  \label{eqvc}
\end{equation}
respectively, where $\tilde{\tau}_i = \sum_{j=1}^i \tau_j$ (with $\tau_0 \equiv 0$), $a_{i}$ and $A_i$ are  integration constants to be determined from the boundary conditions and  $F_i(t)$ can be viewed as a ``time integrated force'', which is
related to the external forces through the expression
\begin{equation}
 F_i(t)=\int_{\tilde{\tau}_{i-1}}^t e^{\gamma_i t^\prime}f_i(t^\prime)dt^\prime.
\end{equation} 
Here, the variable $t$ is interpreted as the time modulus the period $\tau = \tilde{\tau}_N$.

Since the probability distribution is continuous, the conditions 
$\langle v_i\rangle(\tau_i)=\langle v_{i+1}\rangle(\tau_i)$ and  $b_i(\tau_i)=b_{i+1}(\tau_i)$ must hold for $i=1, \ldots N-1$. In addition, the steady state condition (periodicity) implies that $\langle v_1\rangle(0)=\langle v_{N}\rangle(\tau)$ and $b_1(0)=b_{N}(\tau)$. Hence, the $a_i$ and $A_i$ can be determined as the solution of two uncoupled linear systems of $N$ equations each.
Here  we shall focus on the case of $N=2$ reservoirs -- the simplest
case for tackling the efficiency of a thermal engine, in which the interaction with the first and second reservoirs occur during $\tau_1$ and
$\tau_2=\tau-\tau_1$, respectively. For simplicity, from now on, we consider that the viscous constant are equal $\gamma_1=\gamma_2=\gamma$. Therefore, the average velocities and their variances are 
\begin{eqnarray}
  \label{eq:avvel}
  \langle v_1 \rangle (t)&=&\frac{\left(e^{\gamma  \tau }-1\right) F_1(t)+F_1(\tau_1)+F_2(\tau)}{e^{\gamma  t}\left (e^{\gamma  \tau }-1\right)}, \\
\langle v_2 \rangle (t)&=& \frac{e^{\gamma  \tau } F_1(\tau_1)+\left(e^{\gamma  \tau }-1\right) F_2(t)+F_2(\tau)}{e^{\gamma t} \left(e^{\gamma  \tau }-1\right)}, \nonumber
\end{eqnarray}
and
\begin{eqnarray}
  \label{b1}
  b_1(t) &=& -\frac{(\Gamma_{1}-\Gamma_{2}) \left(1-e^ {-2\gamma\tau_2}\right)}{2 \gamma \left(1-e^ {-2 \gamma \tau}\right)}e^{-2\gamma t}+\frac{\Gamma_{1}}{2\gamma},\\
  b_2(t) &=&\frac{(\Gamma_{1}-\Gamma_{2}) \left(1-e^ {-2\gamma\tau_1}\right)}{2 \gamma \left(1-e^ {-2 \gamma \tau}\right)} e^{-2\gamma(t-\tau_1)}+\frac{\Gamma_{2}}{2\gamma}, \nonumber
\end{eqnarray}
respectively.
The expressions for $\langle v_1 \rangle(t)$ and $b_1(t)$ hold for $0\le t \leq  \tau_1$, while the expressions for $\langle v_2 \rangle(t)$ and $b_2(t)$ are valid for $\tau_1 \le t \leq \tau$. It is worth pointing out that the particle will be exposed to the contact with the reservoir 1 and force $f_1(t)$ for a longer (shorter) time than with reservoir 2 and force $f_2(t)$ if $\tau_1 \ge \tau_2 \;(\tau_1 \le \tau_2)$. Furthermore, while the average velocities $\langle v_i \rangle (t)$ depend on the external force (but not on the temperature of the reservoirs), its variances $b_i(t)$ depend on the temperatures (but not on the external forces).

Having  the expressions for the mean velocities and variances, thermodynamic quantities of interest can be directly obtained.
The average work in each part of the cycle is given by
\begin{eqnarray}
  \overline{\dot{W}}_1 &=& \frac{1}{\tau} \int_{0}^{\tau_1} \langle v_1 \rangle (t) f_1(t) dt, \label{eq:w1}\\
  \overline{\dot{W}}_2 &=& \frac{1}{\tau} \int_{\tau_1}^{\tau} \langle v_2 \rangle (t) f_2(t) dt. \label{eq:w2}
 \end{eqnarray}
Using Eq.~(\ref{eq:avvel}) and expressing each external force 
as $f_i(t)=X_i g_i(t)$, with  $X_i$ and $g_i(t)$ denoting
 force strength and its driving, respectively, we finally
arrive at the following expressions:
\begin{widetext}
  \begin{eqnarray}
     \label{l11w}
  \overline{\dot{W}}_1 &=& -\frac{m}{\tau  \left(e^{\gamma  \tau }-1\right)} \left[X_1^2 \left(\left(e^{\gamma  \tau}-1\right) \int_0^{\tau_1}   g_1(t) e^{-\gamma t}\, dt\int_0^t g_1(t^\prime) e^{\gamma t^\prime} \, dt^\prime +\int_0^{\tau_1} g_1(t) e^{-\gamma t} \, dt \int_0^{\tau_1} g_1(t^\prime) e^{\gamma  t^\prime} \, dt^\prime\right) \right . \nonumber\\
  & + & X_1 X_2 \left . \int_0^{\tau_1} g_1(t) e^{-\gamma t} \, dt \int_{\tau_1}^{\tau } g_2(t^\prime) e^{\gamma  t^\prime} \, dt^\prime\right],\\ 
  \overline{\dot{W}}_2 &=& -\frac{m}{\tau  \left(e^{\gamma  \tau }-1\right)} \left[X_2^2 \left (\int_{\tau_1}^{\tau } g_2(t) e^{-\gamma  t} \, dt \int_{\tau_1}^{\tau } g_2(t^\prime) e^{\gamma  t^\prime} \, dt^\prime+\left(e^{\gamma  \tau }-1\right) \int_{\tau_1}^{\tau } g_2(t) e^{-\gamma  t} dt\int_{\tau_1}^t g_2(t^\prime) e^{\gamma  t^\prime} \, dt^\prime \,  \right) \right . \nonumber \\
    &  + & \left . X_1 X_2 e^{\gamma  \tau }  \int_{\tau_1}^{\tau } g_2(t) e^{-\gamma  t} \, dt\int_0^{\tau_1} g_1(t^\prime) e^{\gamma  t^\prime} \, dt^\prime\right].
  \label{l22w}
\end{eqnarray}
\end{widetext}
The expressions above, Eqs.~(\ref{l11w}) and~(\ref{l22w}), are exact and are valid for any kind of
drivings $g_1(t)$ and $g_2(t)$  and stage duration $\tau_1$ and $\tau_2$. 
Usually, in the linear regime, $\overline{\dot{W}}_i$ is written as the product of a flux ${\cal J}_{i}=L_{ii}X_i+L_{ij}X_j$ by a force $X_i$, that is, $\overline{\dot{W}}_i=- \kB T_i {\cal J}_{i}X_i$. Since in the present case $\overline{\dot{W}}_i$ is always bilinear in the forces $X_i$, such expression is also valid even far from the linear regime. Thus, the Onsager coefficients $L_{ij}$ may be written as,
\begin{widetext}
  \begin{equation}
    \begin{split}
   L_{11}&=  \frac{2\gamma}{\Gamma_1\tau  \left(e^{\gamma  \tau }-1\right)} \left[\left(e^{\gamma  \tau}-1\right) \int_0^{\tau_1}g_1(t) e^{-\gamma  t}dt\int_0^t  g_1(t^\prime) e^{\gamma t^\prime} \, dt^\prime \, +\int_0^{\tau_1} g_1(t) e^{-\gamma t} \, dt \int_0^{\tau_1} g_1(t^\prime) e^{\gamma  t^\prime} \, dt^\prime\right],\\
L_{22}&=\frac{2\gamma}{\Gamma_2\tau  \left(e^{\gamma  \tau }-1\right)}\left [\int_{\tau_1}^{\tau } g_2(t) e^{-\gamma  t} \, dt \int_{\tau_1}^{\tau } g_2(t^\prime) e^{\gamma  t^\prime} \, dt^\prime+\left(e^{\gamma  \tau }-1\right) \int_{\tau_1}^{\tau } g_2(t) e^{-\gamma  t}  dt \int_{\tau_1}^t g_2(t^\prime) e^{\gamma  t^\prime} \, dt^\prime \,\right ],\\
  L_{12}&=\frac{2\gamma}{\Gamma_1\tau  \left(e^{\gamma  \tau }-1\right)} \int_0^{\tau_1} g_1(t) e^{-\gamma t} \, dt \int_{\tau_1}^{\tau } g_2(t^\prime) e^{\gamma  t^\prime} \, dt^\prime,\\
  L_{21}&=\frac{2\gamma e^{\gamma  \tau }}{\Gamma_2\tau  \left(e^{\gamma  \tau }-1\right)}   \int_0^{\tau_1} g_1(t^\prime) e^{\gamma  t^\prime} \, dt^\prime \int_{\tau_1}^{\tau } g_2(t) e^{-\gamma  t} \, dt.
  \label{l21ew}
    \end{split}
\end{equation}
\end{widetext}
Reciprocal relations  are verified
  as follows: Since forces $f_1(t)$ and $f_2(t)$ solely act from 0 to $\tau_1$ and $\tau_1$ to $\tau$,  respectively, both upper and lower  integral limits
  in Eqs.~(\ref{eq:w1}) and Eq.~(\ref{eq:w2})  can be replaced for $\tau$ and
  $0$, respectively and hence  all expressions from Eq.~(\ref{eq:w1}) to Eq.~(\ref{l21ew}) can be evaluated over a complete cycle. By exchanging the indexes $1 \leftrightarrow 2$, we verify that $L_{ij}\leftrightarrow L_{ji}$.

\begin{figure}[!h]
  \centering
  \includegraphics[scale=1]{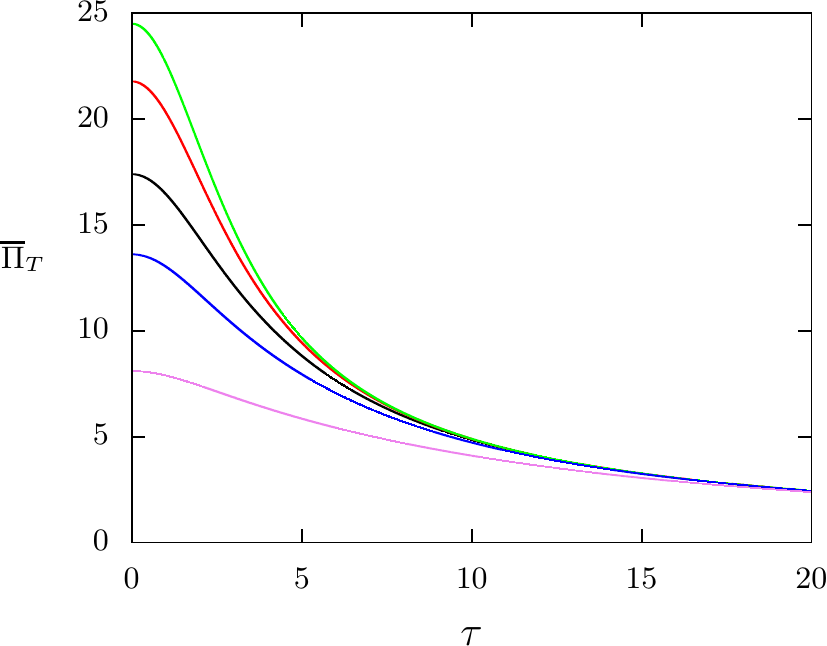}
  \caption{Mean entropy production component ${\overline \Pi}_T$
  as a function of the period $\tau$ for $\gamma=1, \; \Gamma_1=1$ and $\Gamma_2=100$ and distinct asymmetries. From top to bottom: $\kappa=1.0,\;0.5,\;0.3,\;0.2$ and $0.1$.}
\label{fig1}
\end{figure}

Similarly, general expressions can be obtained for the average heat dissipation during the contact of the Brownian particle with each reservoir. Since the heat is closely related to the entropy production rate [see e.g.\ Eq.~(\ref{eqf})], we curb our discussion to the latter quantity. The average entropy production over a complete cycle is then given by,
\begin{equation}
   {\overline\Pi}= \frac{1}{\tau}\left[ \int_{0}^{\tau_1} \Phi_{1} (t)\,dt+ \int_{\tau_1}^{\tau} \Phi_2(t) \, dt\right].
   \label{pibar}
\end{equation}
By inserting Eq. (\ref{eqf}) into Eq. (\ref{pibar}) and using Eq. (\ref{112b}), ${\overline\Pi}$ can be decomposed in two terms:
one associated with the difference of temperature of the reservoirs 
\begin{equation}
   {\overline\Pi_T}= \frac{k_B}{\tau}\left[\frac{2\gamma^2}{\Gamma_1} \int_{0}^{\tau_1} b_{1} (t)\,dt+ \frac{2\gamma^2}{\Gamma_2}\int_{\tau_1}^{\tau} b_2(t) \, dt-\gamma \tau\right],
  \label{ept}
\end{equation}
and the other coming from the external forces 
\begin{equation}
  {\overline\Pi_F}= \frac{k_B}{\tau}\left[\frac{2\gamma^2}{\Gamma_1} \int_{0}^{\tau_1} \langle v_{1}\rangle^2(t)\,dt+ \frac{2\gamma^2}{\Gamma_2}\int_{\tau_1}^{\tau} \langle v_{2}\rangle^2(t) \, dt \right].
  \label{epf}
\end{equation}
Now, from Eqs.~(\ref{b1}) and~(\ref{ept}), one obtains
the general form for $ {\overline \Pi_T}$:
\begin{equation}
   {\overline \Pi_T}=k_B \frac{\Gamma_{1} \Gamma_{2}}{\tau}\frac{\sinh \left(\gamma \tau_1 \right)\sinh \left(\gamma \tau_2 \right)}{\sinh \left(\gamma \tau \right)}\left(\frac{1}{\Gamma_{1}}-\frac{1}{\Gamma_{2}}\right)^{2},
   \label{pit}
\end{equation}
which it is strictly positive (as expected). The
component $(1/\Gamma_{1}-1/\Gamma_{2})$
can be regarded as  the ``thermodynamic force''  $f_\Gamma$
associated with the difference of temperature of the
reservoirs. Particularly, in the linear regime ($\Gamma_2 \simeq \Gamma_1 = \Gamma$),   ${\overline \Pi_T}$  can be conveniently written
down in terms of   Onsager coefficient
${\overline \Pi_T}=L_{\Gamma \Gamma}f_\Gamma$, where $L_{\Gamma \Gamma}$
is given by,
\begin{equation}
   L_{\Gamma \Gamma}=k_B \frac{\Gamma^2}{\tau}\frac{\sinh{(\gamma\tau_1)}\sinh{(\gamma\tau_2)}}{\sinh{(\gamma\tau)}}.
\end{equation} 
Note that $L_{\Gamma \Gamma}$ is strictly positive and it reduces to 
   $k_B \Gamma^2 \tanh{[\frac{\gamma\tau}{2}]}/ 2 \tau$
   for $\tau_1 = \tau_2$ (symmetric case). Further,
it is straightforward to verify that the dissipation term $\Pi_T$
is a monotonous decreasing function of $\tau$ and it is always larger for the symmetric case ($\tau_1 = \tau_2$). Both properties of $\overline{\Pi}_T$ are illustrated in Fig.~\ref{fig1}, where ${\overline \Pi}_T$ is shown as a function of
$\tau$ for various values of the asymmetry parameter $\kappa=\tau_1/\tau_2$ (notice that $\overline{\Pi}_T$ is invariant over the switch of the interaction times $\tau_1 \leftrightarrow \tau_2$ or, equivalently $\kappa \leftrightarrow 1/\kappa$). There is one caveat which concerns the validity of the results of Fig.~\ref{fig1}. Collisional models usually neglects the time for changing the contact between the system and thermal baths. However, if $\tau$ is very small, such approximation can no longer be hold. We  shall assume  along this paper
that $\tau$ is large enough for the collisional approximation to be valid.

The entropy production component coming from external forces also assumes a general (bilinear) form given by,  
\begin{equation}
  {\overline \Pi}_F= {\tilde L}_{11} X_{1}^{2}+({\tilde L}_{12}+{\tilde L}_{21}) X_{1} X_{2}+{\tilde L}_{22}X_{2}^{2}.
  \label{bilinear}
\end{equation}
The coefficients ${\tilde L}_{ij}$'s are shown in the Appendix~\ref{app1}, Eq.~(\ref{l11}).
It should be noticed that Eq.~(\ref{bilinear}) is exact for all force regimen (not only in the linear regime).  For equal temperatures, they
coincide with  Onsager coefficients [Eq.~(\ref{l21ew})].
A detailed analysis for distinct linear regimes (low temperature
difference and/or low forces) is undertaken
in Appendix~\ref{app1}.
Furthermore, since $\tau_2=\tau-\tau_1$, the 
coefficients above fulfill the reciprocal relations
${\tilde L}_{11}\leftrightarrow {\tilde L}_{22}$ and ${\tilde L}_{12}\leftrightarrow {\tilde L}_{21}$ by exchanging $1\leftrightarrow 2$ for the generic drivings $g_i(t)$'s, the interaction times $\tau_i$'s and the temperature of the reservoirs $T_i$'s.

\section{Efficiency}
The optimization of  engines, which converts energy (usually heat or chemical work)
into mechanical work,
constitutes one of the main issues
in thermodynamics, engineering, chemistry and others. Here we exploit
the role of asymmetric contact times between the Brownian particle and the thermal reservoirs as a reliable strategy for optimizing the
machine performance. More specifically,
the amount of energy (heat and work) received by the particle  is  partially converted into output work (or, equivalently, the output power per cycle)
 ${\cal P}={\overline {\dot W_2}}\ge 0$ during the second half stage.
A measure of efficiency is given by the ratio of the amount of
output work to the total energy injected
\begin{equation}
  \eta=-\frac{\cal P}{{\overline {\dot W_1}}+{\overline {\dot Q_i}}},
  \label{eff1}
\end{equation}
where ${\overline {\dot Q_i}}$ is the average
heat extracted from the reservoir $i$  ($i=1$ or $2$ whether the reservoir $1$ or $2$ delivers heat to the Brownian particle), whereas for
the other way round (both reservoirs absorbing energy from the  particle), $\overline{Q}_i$ does not appear in Eq.~(\ref{eff1}), as shall be discussed in Sec.~\ref{equalt}.
Below, we are going to investigate the  machine optimization 
with respect  to the loading force $X_2$ and asymmetry coefficient $\kappa=\tau_1/\tau_2$ for two distinct scenarios: equal and different
temperatures.  

\subsection{Isothermal work-to-work converter}\label{equalt}
Many processes in nature, such as biological systems, operate at
homogeneous (or approximately equal) temperatures, in which an amount of chemical work/energy
is converted into mechanical work and vice-versa (see e.g.\ \cite{liepelt2,altaner}). This highlight the importance of searching for optimized
protocols operating at equal temperatures.
 Here we exploit the present Brownian machine operating
 at equal temperatures, but subject to distinct external forces. From Eqs.~(\ref{112b}) and~(\ref{b1}), it follows  that ${\overline {\dot{Q}_1}} \geq 0$ and ${\overline {\dot{Q}_2}} \geq 0$ and therefore no heat is delivered to the particle. Such engine
reduces to a work-to-work converter: the particle receives input power
${\overline {\dot W_1}}<0$  which  is partially converted
into output power ${\cal P} \ge 0$. From Eq.~(\ref{l21ew}),  the output power
and efficiency can expressed  in term of the Onsager coefficients according to the following expressions:
\begin{equation}
 {\cal P} = \overline{\dot{W}}_2  = - \kB T \left[L_{22}(\kappa) X_2^2 + L_{21} (\kappa)X_1 X_2\right]. 
 \label{eq:power}
\end{equation}
and
\begin{equation}
  \eta=-\frac{L_{21}X_1X_2+L_{22}X_2^2}{L_{11}X_1^2+L_{12}X_1X_2},
  \label{efff}
\end{equation}
Both of them can be   expressed in terms of the ratio  $X_2/X_1$ between forces, the output power being a function of such ratio multiplied by $X_1^2$.
As mentioned previously,   there are three routes to be considered with respect to the engine optimization (holding $X_1$ and $\tau$ fixed):
the time asymmetry optimization (conveniently carried out in
terms of ratio $\kappa=\tau_1/\tau_2$) the output force  $X_2$ optimization; and both optimizations together. We shall analyze all cases in the following subsections.

\subsubsection{Maximization with respect to the asymmetry}

Since the Brownian particle must be in contact with the first reservoir long enough for the injected energy to be larger than the energy dissipated by the viscous force,  for any set of $X_1$ and $X_2$ there is a minimum value $\kappa_m$  for which ${\cal P}\ge 0$. On the other hand, depending on the kind of driving, it can extend up $\kappa \rightarrow \infty$, for which  $L_{21}$ and $L_{22}$ vanishes [see Eq.~(\ref{l21ew})].  

The choice of optimal asymmetries are expected to be dependent of
the quantity chosen to be maximized. Usually, there are two quantities of interest: maximum efficiency or  maximum power output. Starting with the latter case, the optimal asymmetry $\kappa_{MP}$ which maximizes ${\cal P}$ is
 the solution of following equation
\begin{equation}
  \frac{L'_{21}(\kappa_{MP})}{L'_{22}(\kappa_{MP})} = -\frac{X_2}{X_1},
  \label{eq:kappamP}
\end{equation}
where  $L_{ij}'(\kappa)\equiv\partial L_{ij}(\kappa)/\partial \kappa$
and in this section  $L_{ij}$'s (together their derivatives) have been
expressed in terms of $\kappa$ for
specifying which quantity (${\cal P}$ or $\eta$)  has been maximized. In general, Eq.~(\ref{eq:kappamP}) may have more than one solution for each choice of the ratio $X_2/X_1$ and one should be careful to identify the global maximum. However, in the following discussion (as in the examples presented in Section~\ref{models}), we consider the cases which present a single maximum.

Similarly, from Eq.~(\ref{efff}), we obtain the value of the asymmetry that maximizes the efficiency $\kappa_{M\eta}$ from the transcendental equation
\begin{equation}
  \begin{split}
  \Delta_{2212}(\kappa_{M\eta}) &X_2^2 + \Delta_{2111}(\kappa_{M\eta}) X_1^2 \\
  &+ \left[ \Delta_{2211}(\kappa_{M\eta}) + \Delta_{2112}(\kappa_{M\eta}) \right] X_1 X_2 = 0,  
  \end{split}
  \label{eq:kappameta}
\end{equation}
where $\Delta_{ijkl}(\kappa) = L'_{ij}(\kappa) L_{kl}(\kappa) - L'_{kl}(\kappa) L_{ij}(\kappa)$. Although exact,  for a given choice of the drivings $g_i(t)$ and the strengths $X_i$, Eqs.~(\ref{eq:kappamP}) and~(\ref{eq:kappameta}), in general, have to be solved numerically for $\kappa_{MP}$ and $\kappa_{M\eta}$, respectively. After these values are obtained, we can evaluate the power ${\cal P}_{MP,\kappa}$ and efficiency $\eta_{MP,\kappa}$ at maximum power as
\begin{equation}
  {\cal P}_{MP,\kappa}=\frac{\kB T L_{21}'(\kappa_{MP})}{L_{22}'^2(\kappa_{MP})}[L_{21}(\kappa_{MP})L_{22}'(\kappa_{MP})-L_{22}(\kappa_{MP})L_{21}'(\kappa_{MP})]X_1^2,
\end{equation}
and
\begin{equation}
\eta_{MP,\kappa}=\frac{L_{21}'(\kappa_{MP})[L_{22}'(\kappa_{MP})L_{21}(\kappa_{MP})-L_{22}(\kappa_{MP})L_{21}'(\kappa_{MP})]}{L_{22}'(\kappa_{MP})[L_{11}(\kappa_{MP})L_{22}'(\kappa_{MP})-L_{12}(\kappa_{MP})L_{21}'(\kappa_{MP})]}.
\end{equation}

 Analogously, we can write the power at maximum efficiency ${\cal P}_{M\eta,\kappa}$ and maximum efficiency $\eta_{M\eta,\kappa}$  as
\begin{equation}
  {\cal P}_{M\eta,\kappa}= -\kB T \left[L_{22}(\kappa_{M\eta}) X_2^2 + L_{21} (\kappa_{M\eta})X_1 X_2\right], 
\end{equation}
and
\begin{equation}
  \eta_{M\eta,\kappa}=- \frac{L_{22}(\kappa_{M\eta}) X_2^2 + L_{21} (\kappa_{M\eta})X_1 X_2}{L_{11}(\kappa_{M\eta}) X_1^2 + L_{12} (\kappa_{M\eta})X_1 X_2},
\end{equation}
respectively.
In Sec~\ref{models}, we will exemplify our exact
expressions for maximum efficiencies and powers for two kinds of drivings.

\subsubsection{Maximization with respect to the output force\label{sec:maxforce}}
For given asymmetry and drivings, the Onsager coefficients are constant. Hence, the maximization of the output power and the efficiency turn out to be similar to the approach  from Refs.~\cite{karel2016, noa2020thermodynamics}. Below, we recast the main  results.

As previously, the engine regime (${\cal P}>0$) also imposes boundaries to optimization with respect to the force strength. Here, the output force $X_2$  must lie in the interval $X_m\le X_2\le 0$, where $X_m=-L_{21}X_1/L_{22}$. In general, $X_m$ is different from the value of the output force that minimizes the entropy production $X_{2mS}$ (for $X_1$ and $\kappa$ constants). However, they coincide $X_m = X_{2mS}$ for symmetric Onsager coefficients $L_{12}=L_{21}$.
Similarly to the previous subsection, the optimization can be performed to ensure maximum power ${\cal P}_{MP,X_2}$  (with efficiency ${\eta}_{MP,X_2}$)
or maximum efficiency ${\eta}_{M\eta,X_2}$   (with power ${\cal P}_{M\eta,X_2}$), by adjusting
the output forces  to optimal values $X_{2MP}$ and $X_{2M\eta}$, respectively. These optimal output forces can be expressed in terms of the Onsager coefficients as
\begin{equation}
  X_{2M\eta}=\frac{L_{11} }{L_{12} }\left(-1+ \sqrt{1-\frac{L_{12} L_{21}}{L_{11} L_{22}}}\right)X_1,
  \label{eq:x2meta}
\end{equation}
and
\begin{equation}
  X_{2MP}=-\frac{1}{2}\frac{L_{21}}{L_{22}}X_1,
  \label{x2mp}
\end{equation}
respectively. 
Hence, the  maximum efficiency $\eta_{M\eta,X_2}$ and the efficiency at maximum power $\eta_{MP,X_2}$ are given by, 
\begin{equation}
  \eta_{M\eta,X_2}=-\frac{L_{21}}{L_{12}}+\frac{2L_{11}L_{22}}{L_{12}^2}\left(1-\sqrt{1-\frac{L_{12} L_{21}}{L_{11} L_{22}}}\right),
  \label{etame}
\end{equation}
and
\begin{equation}
  \eta_{MP,X_2}=\frac{L_{21}^2}{4 L_{11}L_{22}-2L_{12}L_{21}},
\label{etamp}
\end{equation}
while the  power at maximum efficiency ${\cal P}_{M\eta,X_2}$
and the maximum power ${\cal P}_{MP,X_2}$ can obtained by inserting $X_{2M\eta}$ or  $X_{2MP}$ into the expression for ${\cal P}$. In fact, these quantities are not independent of each other, instead they are related as
\begin{equation}
  \eta_{MP,X_2}=\frac{P_{MP,X_2}}{2P_{MP,X_2}-P_{M\eta,X_2}}\eta_{M\eta,X_2}.
\end{equation}
Furthermore, for symmetric Onsager coefficients $L_{12}=L_{21}$, there two additional simple relations given by,
\begin{equation}
  \eta_{MP,X_2}=\frac{\eta_{M\eta,X_2}}{1+\eta^2_{M\eta,X_2}} \qquad {\rm and} \qquad \frac{{\cal P}_{M\eta,X_2}}{{\cal P}_{MP,X_2}}=1-\eta^2_{M\eta,X_2}.
  \label{opti}
\end{equation}
As shown in Appendix~\ref{sec:constforce},  $L_{12} = L_{21}$ for constant drivings for any value of $\kappa$. Conversely,  they are in general different ($L_{12} \neq L_{21}$) for linear drivings (see Appendix~\ref{sec:linforce}). For the symmetric time case ($\kappa=1$), however, the equality holds also for linear drivings \cite{noa2020thermodynamics}.

 \subsubsection{Constant and linear drivings\label{models}}
 In order to access the advantages of the asymmetry in the time spent by the Brownian particle in contact with each reservoir, we consider two different driving models. In the first model, the drivings are constant and the external forces can be written as
\begin{eqnarray}\label{eq:constforce1}
  f_1(t) & = & X_1, \quad \text{for} \; 0\le t <\tau_1\\
  f_2(t) & = & X_2, \quad \text{for} \; \tau_1\le t <\tau.
  \label{eq:constforce2}
\end{eqnarray}
In Appendix~\ref{sec:constforce}, we present explicit expressions for the average velocities $\langle v_{i}\rangle(t)$ and  Onsager coefficients $L_{ij}$ (which coincides with the coefficients $\tilde{L}_{ij}$  for isothermal reservoirs). The second class of Brownian engines deals with 
 drivings evolving  linearly in time and given by
the following expressions
\begin{eqnarray} \label{eq:linforce1}
 f_1(t) &=& X_{1} \gamma t, \quad \text{for} \; 0\le t <\tau_1 \\
 f_2(t) &=& X_{2} \gamma (t-\tau_1), \quad \text{for} \; \tau_1\le t <\tau.
 \label{eq:linforce2}
\end{eqnarray}
The main expressions for such case are listed in Appendix~\ref{sec:linforce}.
Figs.~\ref{equal} and~\ref{equal2} depict typical plots of the efficiency and power output for both force models as a function of the output force $X_2$ and asymmetry $\kappa$, respectively.
\begin{figure}
  \includegraphics[scale=0.8]{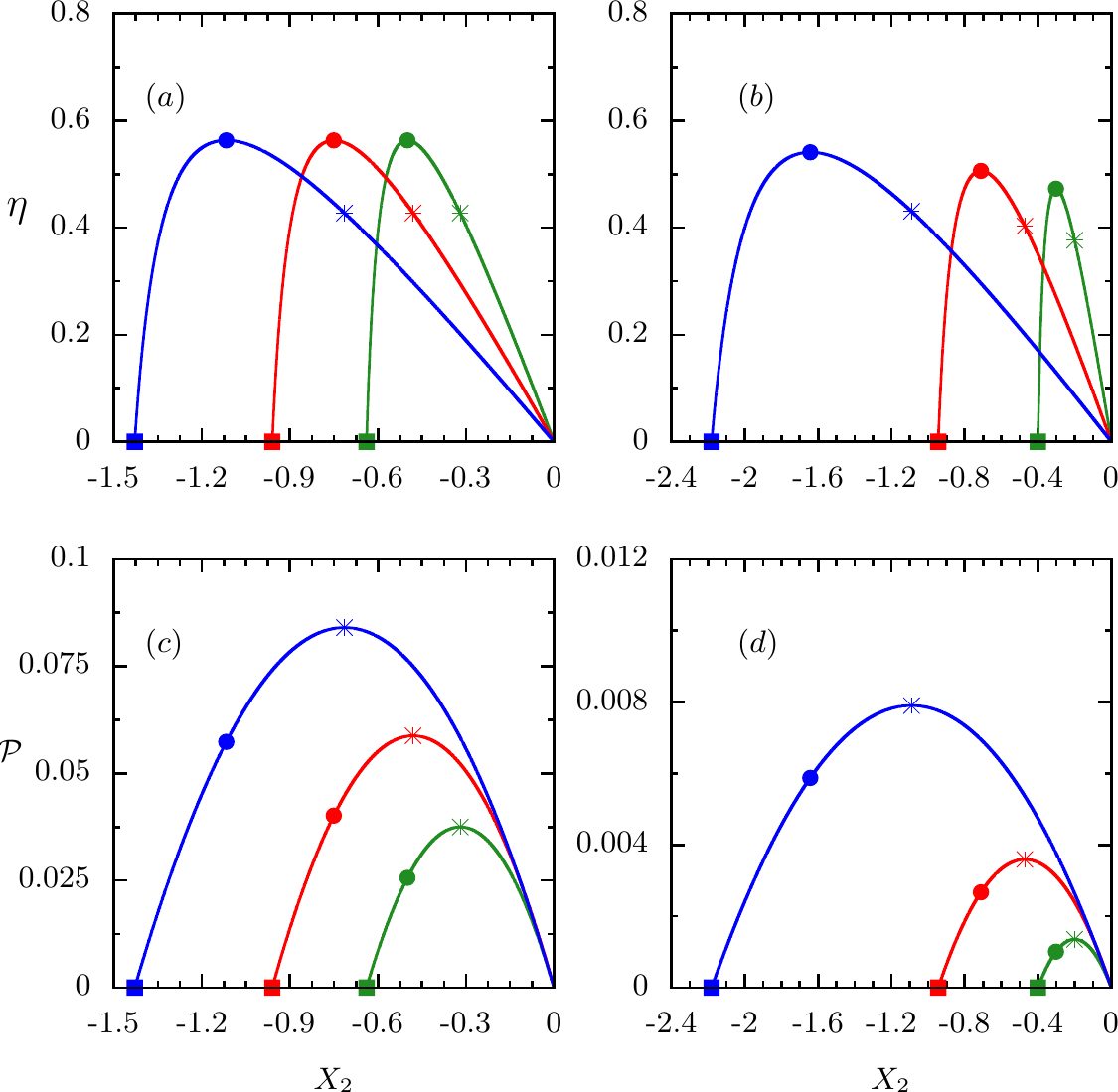}
  \caption{Efficiency [panels $(a)$ and $(b)$] and output power [panels $(c)$ and $(d)$] (averaged over one period) for the isothermal work-to-work converter with $X_1=1$, $\tau=1$, $\gamma=1$ and different asymmetries $\kappa$ (from left to right, $\kappa = 1.50, \; 1.00$ and $0.67$). Panels $(a)$ and $(c)$ depict the results for  constant drivings, whereas $(b)$ and $(d)$ for the linear drivings one. In all panels, squares, circles and stars denote
    $X_{2mS},X_{2M\eta}$ and $X_{2MP}$, respectively.}
    \label{equal}
\end{figure}

As discussed above, the engine regime operates for $X_{2m} < X_2 < 0$. An
immediate advantage of the time asymmetry concerns the minimum output forces $X_{2m}$ which decreases with $\kappa$,
implying that  the engine regime interval increases with the asymmetry (see Fig.~\ref{equal}). Such trend is consistent
  with the absorption of energy (average work rate ${{\overline {\dot W_1}}}$) for longer and longer time as $\kappa$ increases. Furthermore, the minimum entropy production (represented by the squares in the figure) coincides with the minimum loading force (vanishing power output and efficiency) for constant
  drivings, but not for the linear case (although $X_{2mS}$ is close to $X_{m}$). 

  The maximum efficiencies are almost constant for the constant force model [Fig.~\ref{equal}$(a)$] and slightly increase with $\kappa$ [Fig.~\ref{equal}$(b)$] for the linear force model. However, for small $|X_2|$, the efficiency is larger for the smaller values of $\kappa$. The effect of the time asymmetry for the output power is more pronounced. For both force models, the maximum power output clearly increases with $\kappa$. 
\begin{figure}
\centering
\includegraphics[scale=0.7]{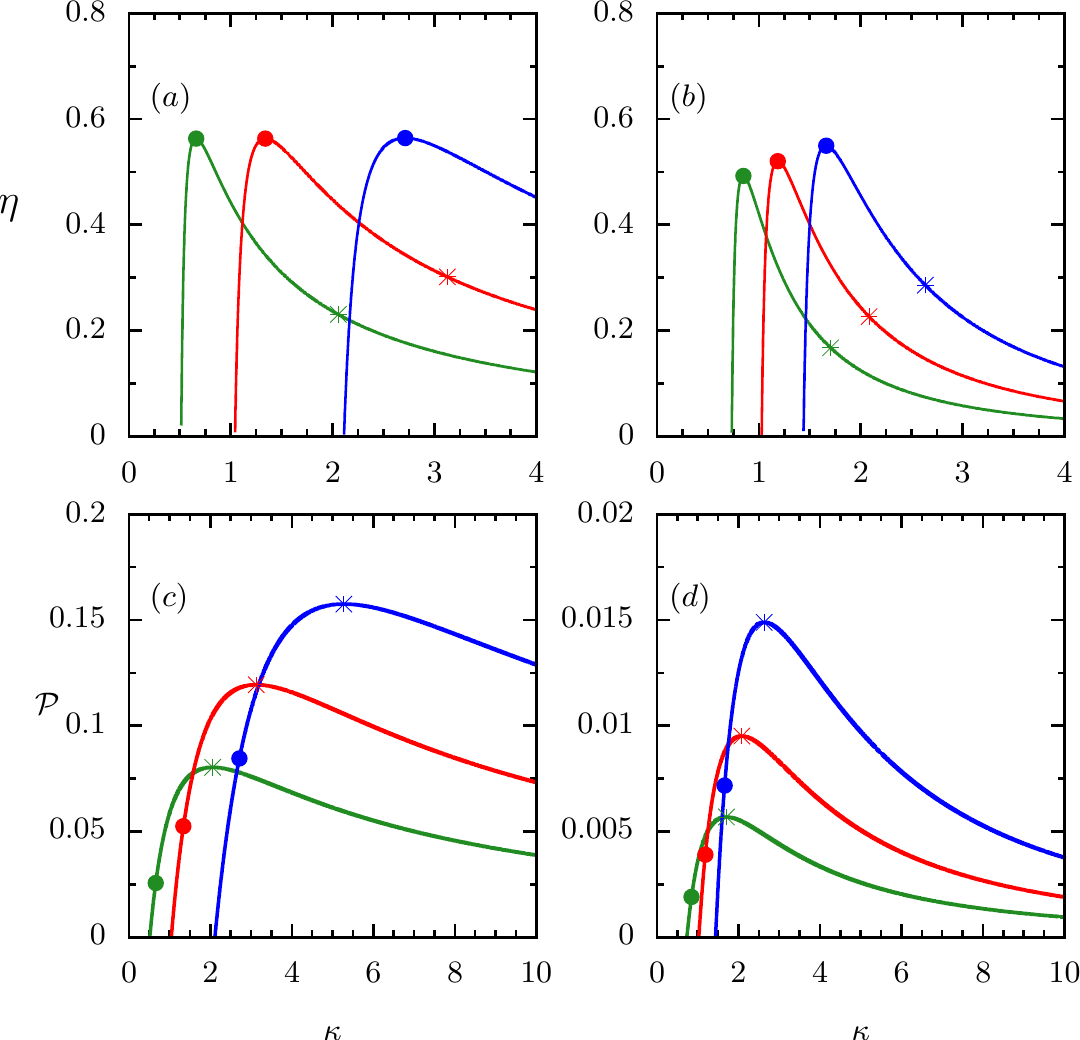}
\caption{Efficiency [panels $(a)$ and $(b)$] and output power [panels $(c)$ and $(d)$] (averaged over one period) for the isothermal work-to-work converter with $X_1=1$, $\tau=1$, $\gamma=1$ and different values of $X_2$ (from left to right, $X_2=-0.5,-1.0$ and $-2.0$). Panels $(a)$ and $(c)$ depict the main results for the constant drivings model while $(b)$ and $(d)$ the linear drivings one. In all panels,  circles and stars denote $\kappa_{2M\eta}$ and $\kappa_{2MP}$, respectively. For such set of parameters, the associate $\kappa_{2mS}$'s are out of
the engine regime.}
    \label{equal2}
\end{figure}

\begin{figure}
    \includegraphics[scale=0.5]{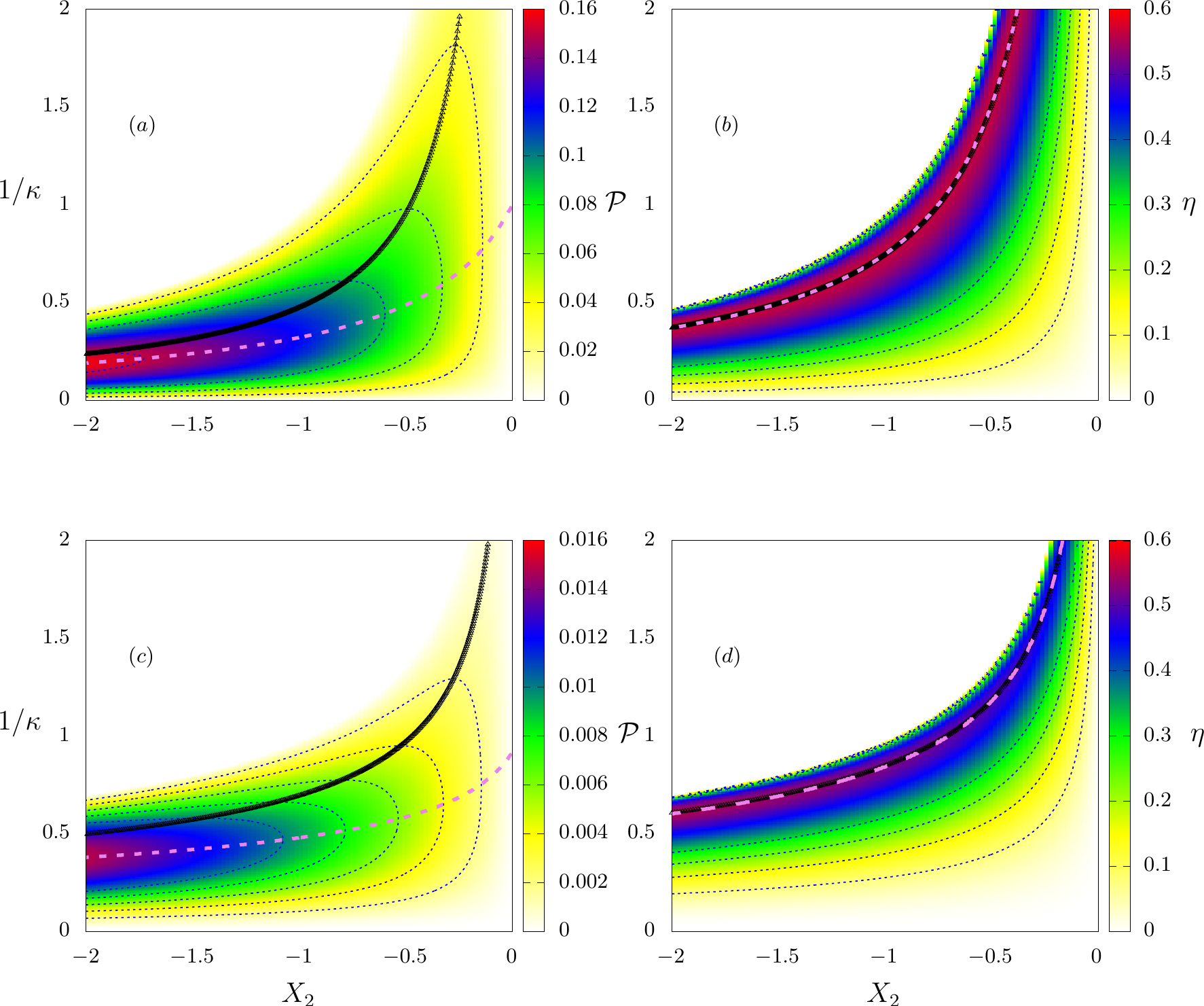}
      \caption{For the isothermal work-to-work converter, the output power (left panels) and efficiency (right panels) for the constant [$(a)$ and $(b)$] and linear [$(c)$ and $(d)$] drivings models as a function of the inverse of the asymmetry parameter $\kappa$ and loading forces $X_{2}$. 
Dotted lines represent constant value loci, dashed and full lines represent maximization with respect to $\kappa$ and $X_2$, respectively. Parameters:  $\tau=1, \; \gamma=1, \; X_1=1$.}
      \label{powere}
\end{figure}

Fig.~\ref{powere} depicts, for constant and linear drivings, a  heat map for the power output and
efficiency as a function of both the asymmetry and loading forces. For aesthetic reasons, they have been expressed in terms $1/\kappa$ (instead of $\kappa$) in the vertical axis.
Noteworthy, the maximum efficiency curves, represented by the dashed (full) line for the maximization with respect to $\kappa$ (loading force), are close to each other. Consequently, the choice of the parameter to maximize the efficiency is not important for both models presented here. Moreover, as previously discussed, the maximum efficiency is almost constant for the constant drivings model, but increases with $\kappa$ for the linear drivings one. 
  In contrast 
  to the maximum efficiencies,  maximum power curves   (panels~\ref{powere}$(a)$ and~\ref{powere}$(c)$ for constant and linear drivings, respectively) present rather different behaviors depending on the optimization parameter. The ${\cal P}_{MP,\kappa}$
  curves  (dashed lines) always lie above the ${\cal P}_{MP,X_2}$ (full lines) ones and they approach each other as $\kappa \rightarrow \infty$.
  Finally, it is worth pointing out that  while both drivings provide  similar efficiencies,  the constant driving case is clearly more advantageous than the linear  one in terms of the output power.
\subsubsection{Simultaneous maximization of the asymmetry and the force\label{sec:maxboth}}
One  may  also raise the relevant issue of  maximizing
the  power output and efficiency with respect to the asymmetry and output force strength simultaneously. Although this is not possible in some cases (as explained below), we will proceed presenting the framework assuming that such maximization is possible. As before, we shall restrict the analysis to drivings  presenting a single physical solution for Eqs.~(\ref{eq:kappamP}) and~(\ref{eq:kappameta}). If this is not the case, each maximum of these equations should be analyzed individually to assert which is the global maximum in each case. 

Under the assumption above, the maximum power output must satisfy simultaneously Eqs.~(\ref{eq:kappamP}) and~(\ref{x2mp}), that is, we must find the optimal value of the asymmetry $\kappa_{MP}^*$ which satisfy the following condition:
\begin{equation}
  \frac{L'_{21}(\kappa_{MP}^*)}{L'_{22}(\kappa_{MP}^*)} = \frac{1}{2}\frac{L_{21}(\kappa_{MP}^*)}{L_{22}(\kappa_{MP}^*)}.
\end{equation}
Once the optimal asymmetry $\kappa_{MP}^*$ is obtained, the optimal force $X_{2MP}^*$ is calculated from Eq.~(\ref{x2mp}) and given by
\begin{equation}
  X_{2MP}^* = -\frac{1}{2}\frac{L_{21}(\kappa_{MP}^*)}{L_{22}(\kappa_{MP}^*)} X_1.
\end{equation}
Graphically, the condition above is precisely the crossing point between
lines for which the power (or efficiency) is maximized with
respect to $X_2$ and $\kappa$.
However, in some cases, (as illustrated by the constant and linear drivings presented above)
these two lines do not cross at all. The physical reason is that
the power output keeps growing as $\kappa \rightarrow \infty$
(with an appropriate choice of a value of $X_2$ for each $\kappa$).
In other words, for such models, it is advantageous to apply a very
large output force (in modulus) for a short period. Conversely, if
the force model involves a rapidly decaying input driving $g_1 (t)$
and growing output driving $g_2(t)$, an optimal output power may be
found. In such case, the power and efficiency at maximum power are
readily evaluated as
\begin{equation}
  {\cal P}^* =  \frac{\kB T}{4} \frac{L^2_{21}(\kappa_{MP}^*)}{L_{22}(\kappa_{MP}^*)} X_1^2, \label{eq:powboth}
\end{equation}
and
\begin{equation}
  \eta^*=\frac{L^2_{21}(\kappa_{MP}^*)}{4 L_{11}(\kappa_{MP}^*)L_{22}(\kappa_{MP}^*)-2 L_{21}(\kappa_{MP}^*)L_{12}(\kappa_{MP}^*)}.
\end{equation}
Thereby, the optimal output power increases quadratically with the input force while the efficiency is completely determined by the driving force model. It is noteworthy that, despite the apparent temperature dependency of the power output in Eq.~(\ref{eq:powboth}), the temperature cancels out when we use the expressions for the Onsager coefficients [see e.g.\ Eq.~(\ref{l21ew})].
Similar expressions can be obtained for the  simultaneous
maximization of efficiency [by equaling
the ratio $X_2/X_1$ from Eqs.~(\ref{eq:kappameta}) and~(\ref{eq:x2meta})].
Since expressions are more involved,  we abstain to present them here.
In order to illustrate the previous ideas,  we consider an exponential
driving given by
\begin{eqnarray}\label{eq:expdriv1}
  f_1(t) & = & X_1 e^{-9\gamma t}, \quad \text{for} \; 0\le t <\tau_1\\
  f_2(t) & = & X_2 e^{\gamma (t-\tau_1)}, \quad \text{for} \; \tau_1\le t <\tau.\label{eq:expdriv2}
\end{eqnarray} 

\begin{figure}
    \includegraphics[scale=0.5]{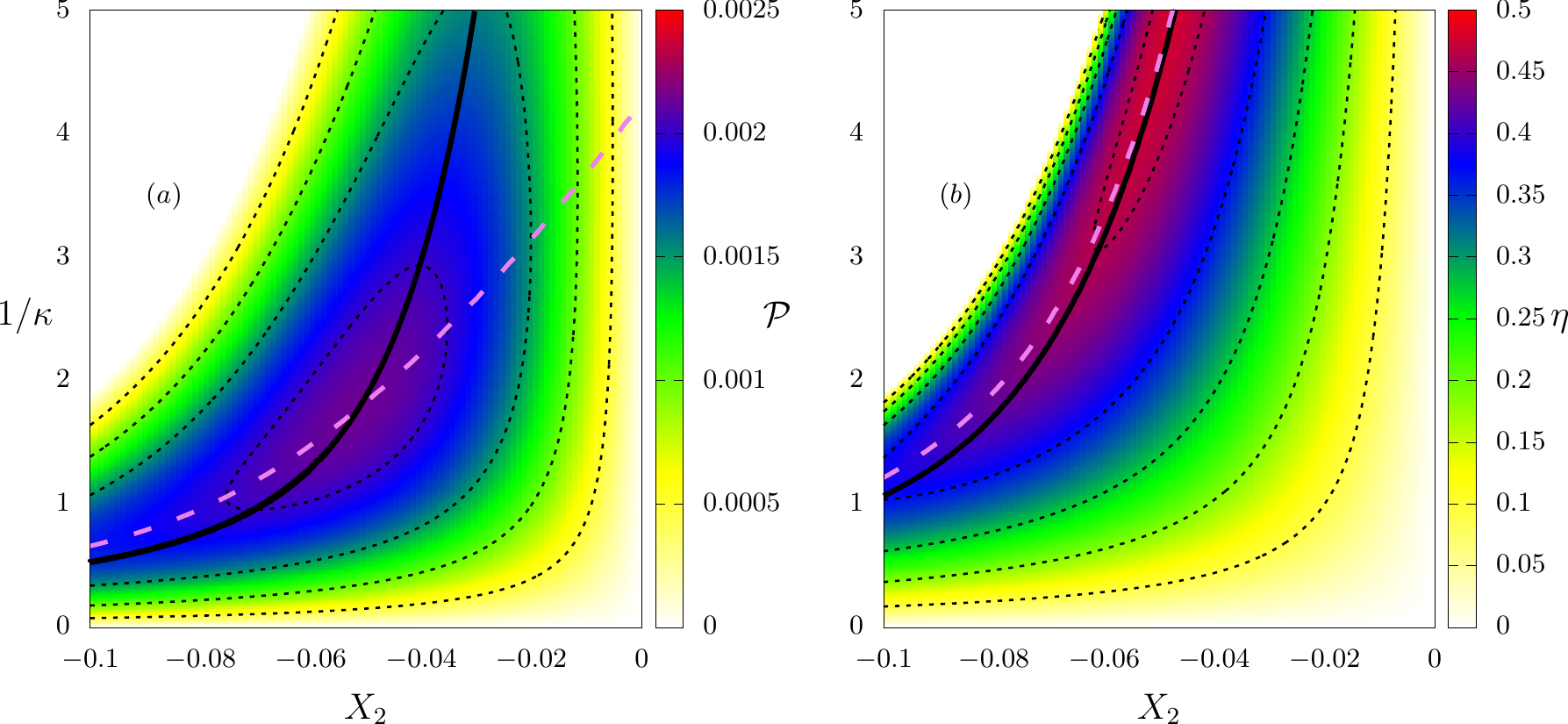}
    \caption{For the exponential driving, depiction of  output power $(a)$
      and efficiency $(b)$ versus the inverse of the asymmetry coefficient $\kappa$ and the output force $X_{2}$ for $\tau=1, \;\gamma=1$ and $X_1=1$. Dotted lines represent constant value loci, dashed and solid lines represent maximization with respect to $\kappa$ and  $X_2$, respectively.}
      \label{heatmapexp}
\end{figure}

Figs.~\ref{heatmapexp} $(a)$ and  $(b)$ depict, for above exponential drivings, the heat maps of the output power and efficiency  as  functions of $\kappa$ and $X_2$, respectively. Contrasting to the previous models, the crossing between maximum power lines are evident for the exponential drivings model above and thereby the global optimization is possible. Although for the exponential model given by Eqs.~(\ref{eq:expdriv1}) and~(\ref{eq:expdriv2}) the crossing between maximum efficiency curves is absent, it does appear for other exponential drivings choices (e.g.\  for $f_1(t)=X_1e^{-7\gamma t}$ and
  $f_2(t)=X_2e^{3\gamma (t-\tau_1)}$) and  follow theoretical prescription above.
 
  \subsection{Thermal engine}\label{thermal}
In this section, we derive general findings for thermal engines in which  the particle  is also exposed to distinct thermal baths  in each stage.
Although the power output ${\cal P}$
is the same as before (it does not depend on the temperatures), the efficiency may change because of the appearance of heat flow. Hence, in addition to the input energy received as work, the engine may also receive energy from the hot reservoir. Consequently, the maximization of power output with respect to the output force $X_{2MP}$ or the asymmetry $\kappa_{MP}$ is the same as before, but the corresponding efficiencies may differ (if ${\overline {\dot Q}_1}<0$ or ${\overline {\dot Q}_2}<0$) from such case, following Eq.~(\ref{eff1}) instead. Anyhow, the efficiency of the engine for reservoirs with different temperatures is always smaller or equal than for isothermal reservoirs.

From Eq.~(\ref{112b}), the average heat dissipated by the Brownian particle per cycle while in contact with the $i$-reservoir ${\overline {\dot Q}_i}$ can be obtained as
\begin{eqnarray}
  {\overline {\dot Q}_1}&=&\frac{m\gamma}{\tau}\left[\int_{0}^{\tau_1} \langle v_1\rangle^2dt-C(\tau_1)(\Gamma_1-\Gamma_2)\right],\label{he}\\
  {\overline {\dot Q}_2}&=&\frac{m\gamma}{\tau}\left[\int_{\tau_1}^{\tau} \langle v_2\rangle^2dt+C(\tau_1)(\Gamma_1-\Gamma_2)\right],\label{he2}
\end{eqnarray}
where $C(\tau_1)=\csch(\gamma \tau) \sinh (\gamma \tau_1) \sinh (\gamma \tau_2)/2\gamma^2$ is strictly positive. Therefore, since the first term on the right-hand side of Eqs.~(\ref{he}) and~(\ref{he2}) are positive, heat always flow into the colder reservoir. As about the hot reservoir, the heat may flow from or into the reservoir. For simplicity, we shall restrict our analysis to the case $\Gamma_1 > \Gamma_2$, that is, the first reservoir being the hot one, but
it is worth pointing out that all the discussion below holds valid for $\Gamma_1 < \Gamma_2$ if we analyze Eq.~(\ref{he2}) instead of Eq.~(\ref{he}). 

For $\Gamma_1 > \Gamma_2$, Eq.~(\ref{he}) ensures that heat flows into the system if $\int_{0}^{\tau_1} \langle v_1\rangle^2dt<C(\tau_1)(\Gamma_1-\Gamma_2)$. Physically, this condition is a balance between kinetic energy that flows into the system due to the forces and the dissipation. If $X_1$ is strong enough (or if the difference of temperature of the reservoirs is small enough), energy flows into both reservoirs. Thereby, the engine effectively reduces to an isothermal work-to-work converter, so that the efficiency is still described by Eq.~(\ref{efff}) and all results and
findings from Section~\ref{equalt} regarding the efficiency optimization hold. Otherwise, the inequality above is satisfied and energy flows from the first reservoir into the engine. For $\Gamma_1 < \Gamma_2$, the same energy balance occurs, but we need to assert the positiveness or negativeness of Eq.~(\ref{he2}).

Furthermore, although exact, the achievement of general expressions for optimized efficiencies outside the isothermal work-to-work regime is more cumbersome than the ones obtained for such regime, making a general analysis unfeasible. Nevertheless, the discussion of a simple asymptotic limit is instructive.
If the second term on the right-hand side of Eq.~(\ref{he}) [or Eq.~(\ref{he2})] is the dominant one, $|\Gamma_1 - \Gamma_2|  \gg 1$ and $|{\overline {\dot Q}_1}| \gg |{\overline {\dot W}_2}|$ [or $|{\overline {\dot Q}_2}| \gg |{\overline {\dot W}_2}|$]. Therefore, the efficiency becomes $\eta\approx-{\cal P}/{\overline {\dot Q}_1}$ [or $\eta\approx-{\cal P}/{\overline {\dot Q}_2}$], which maximization, with respect to $X_2$, yields $ X_{2M\eta} \approx X_{2MP}$ and follows Eq. (\ref{x2mp}). Hence, the corresponding  $\eta_{M\eta}$ approaches to the following expression
 \begin{equation}
   \eta_{M\eta, X_2}\approx\frac{T_2}{8 \gamma^2 T_i C(\tau_1)}\frac{L_{21}^2}{L_{22}}\tau X_1^2 \ll 1,
   \label{rflt}
 \end{equation}
 where $T_i$ is the temperature of the hot reservoir. When the hot
 bath is the first reservoir, the fact that the efficiency is small is direct since the factor $T_2/T_1<<1$. However, when the second reservoir is the hotter one, the temperature ratio becomes 1 and the smallness of the efficiency comes from the Onsager coefficients: $L_{21}^2/L_{22} \propto 1/T_2$. It is also worth mentioning that the apparent dependence on the period cancels out because the Onsager coefficients are proportional to $1/\tau$ [see Eq.~(\ref{l21ew})]. Therefore, for high temperature differences, the engine efficiency is very small for any value of the asymmetry.

\begin{figure}
  \centering
  \includegraphics[scale=0.55]{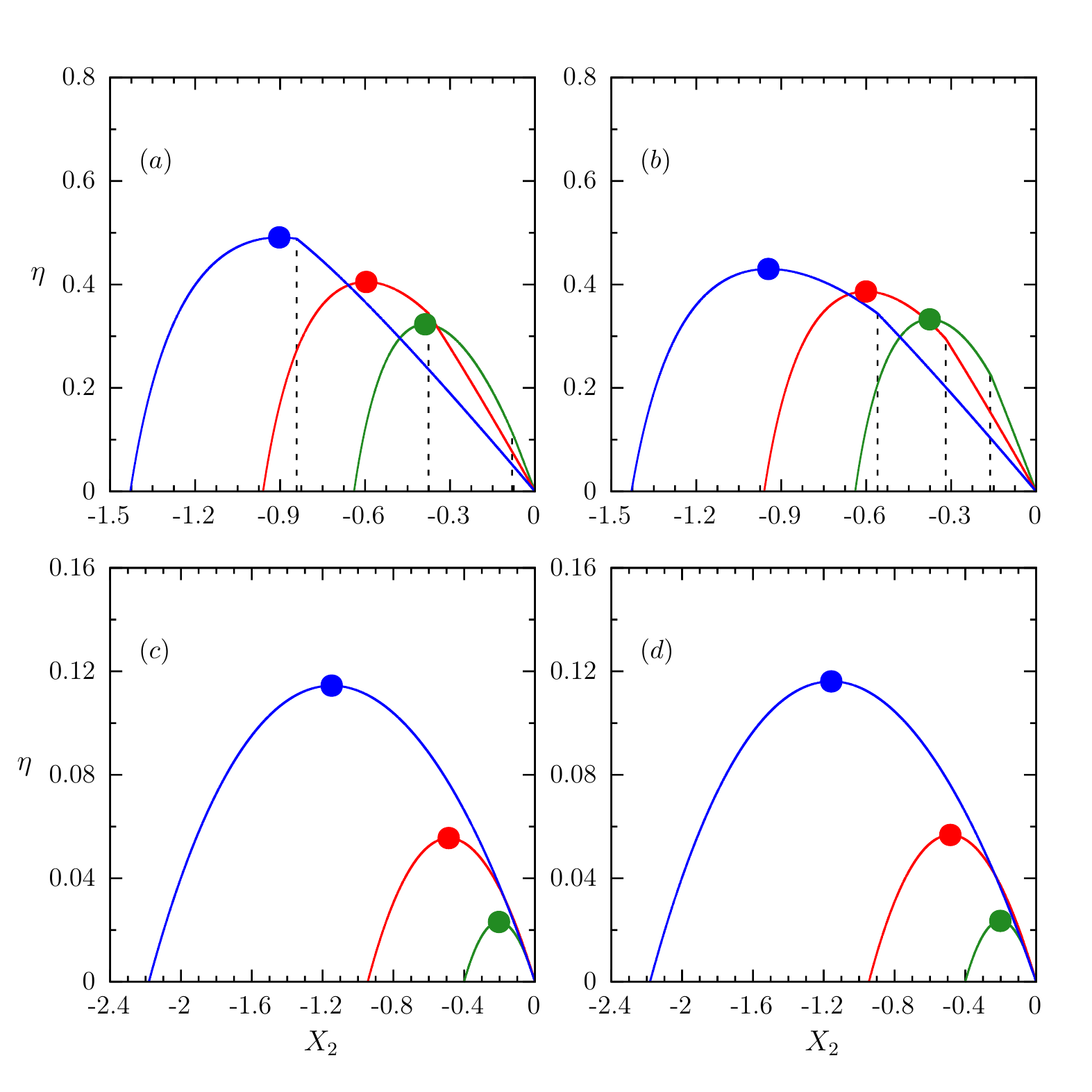}
  \caption{Efficiency as a function of the force strength $X_2$ for the constant [$(a)$ and $(b)$] and linear [$(c)$ and $(d)$] drivings, respectively.
    Parameters: $\tau=1, \; \gamma=1$ and $X_1=1$ and distinct
    temperatures [$\Gamma_1=2.0$ and $\Gamma_2=1.5$ in panels $(a)$ and $(c)$ and $\Gamma_1=1.5$ and $\Gamma_2=2.0$ in panels $(b)$ and $(d)$]. Circles denote maximum efficiencies and their ${X}_{2MP}$'s are the same as in Fig.~\ref{equal}. From left to right, $\kappa = 1.50, \; 1.00$ and $0.67$). Dashed vertical lines stands for the value of $X_2$ for which $\overline{\dot{Q}}_i$ changes sign ($i$ being the index of the hot reservoir).}
    \label{f6}
\end{figure}

In order to illustrate our findings for reservoirs with different temperatures, we consider the constant and linear drivings models presented above. Fig.~\ref{f6} exemplifies, for distinct temperature reservoirs, the efficiency for the same values of $\kappa$ used in Fig.~\ref{equal} for constant [panels $(a)$ and $(b)$] and linear drivings [panels $(c)$ and $(d)$], respectively. In panels $(a)$ and $(c)$ the temperature of the first reservoir is larger than that of the second reservoir, while  panels $(b)$ and $(d)$ depict the other way around.

In accordance with general findings from Sec.~\ref{thermal}, for constant
drivings there are two regimes (the vertical lines in the figure denotes the value of $X_2$ which separates them) for which the heat exchanged between the Brownian particle and the hot reservoir changes sign.
Conversely, they are not present for the linear drivings model -- panels (c) and (d) -- because the heat exchange with the hot reservoir does not change sign for the parameters used in the figures. 
  Since $\langle v_i \rangle^2$ increases with $X_2^2$,
  the  term coming from the difference
  of temperatures in Eq.(\ref{he}) dominates over it when  $|X_2|$ is small
  and hence the machine is less  efficient than the isothermal work-to-work converter. Conversely, for large  $|X_2|$ the engine may become as efficient as the isothermal work-to-work converter if the  exchanged heat with the hot reservoir change sign -- left of the line in panels $(a)$ and $(b)$. Anyhow, by comparing the performance of isothermal with the different  temperature case, we see that
  the decay of efficiency for linear drivings is more pronounced than for  constant drivings.

 As for isothermal reservoirs, the machine performance  always improves as $\kappa$ increases, encompassing not only an extension of its operation regime $X_{2m}$ but it also presents a more pronounced increase of efficiencies, again, more substantial for linear drivings. Moreover, the asymmetry may be used to mitigate the drop in the efficiency produced by the different temperatures of the thermal reservoirs. 

\begin{figure}
    \centering
    \includegraphics[width=8.3cm,height=6.cm]{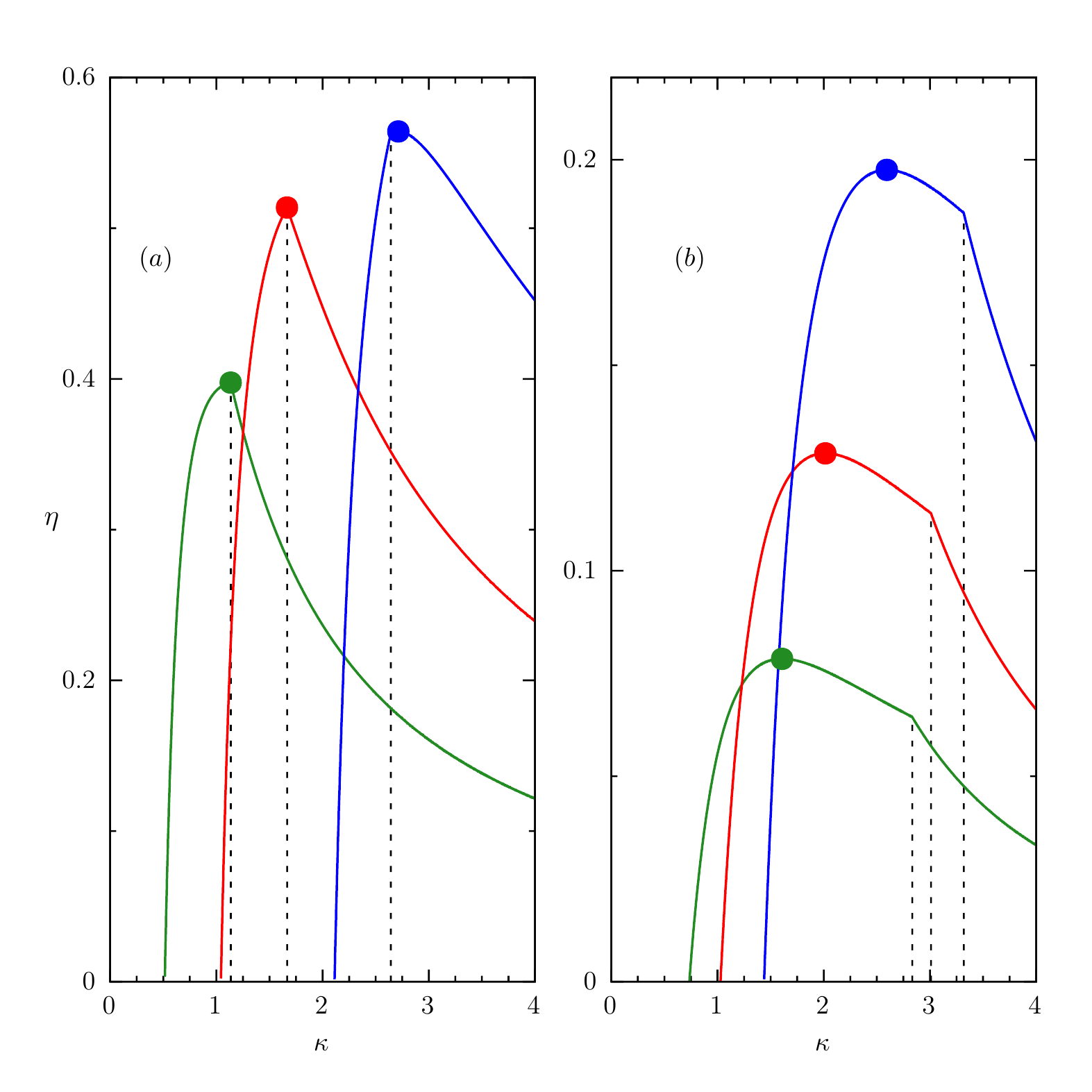}
    \caption{Efficiency versus the time asymmetry $\kappa$ for the $(a)$ constant and $(b)$ and linear  drivings, respectively, for $\tau=1, \; \gamma=1$ and $X_1=1$ and different temperatures [$\Gamma_1=2.0$ and $\Gamma_2=1.5$]. Circles denote maximum efficiencies and their ${X}_{2MP}$'s are the same as in Fig.~\ref{equal2}. From left to right, $X_2=-0.5,-1.0$ and $-2.0$. Dashed vertical lines stands for the value of $\kappa$ for which $\overline{\dot{Q}}_1$ changes sign.  For such set of parameters $k_{2mS}$ are out of the engine regime.}
      \label{figkappa}
\end{figure}

 In Fig.~\ref{figkappa}, we show the efficiency as a function of the asymmetry for various values of $X_2$. Similarly to the previous figure, the vertical lines denote the values of $\kappa$ for which the heat from the hot reservoir change sign and delimits the isothermal work-to-work converter regime. The discussion whether the isothermal work-to-work converter regime lies to the left or right of the vertical lines is not so obvious because both $C(\tau_1)$ and $\langle v_1 \rangle$ depend on the asymmetry. However,  the work-to-work regime lies to the right of the lines, since the function $C(\tau_1)$ reaches its maximum for $\kappa =1 (\tau_1 = \tau/2)$ and the first term on the right-hand side of Eq.~(\ref{he}) is expected to increase giving that
   its limit of integration increases with $\kappa$. 

\begin{figure}
    \centering
    \includegraphics[scale=0.5]{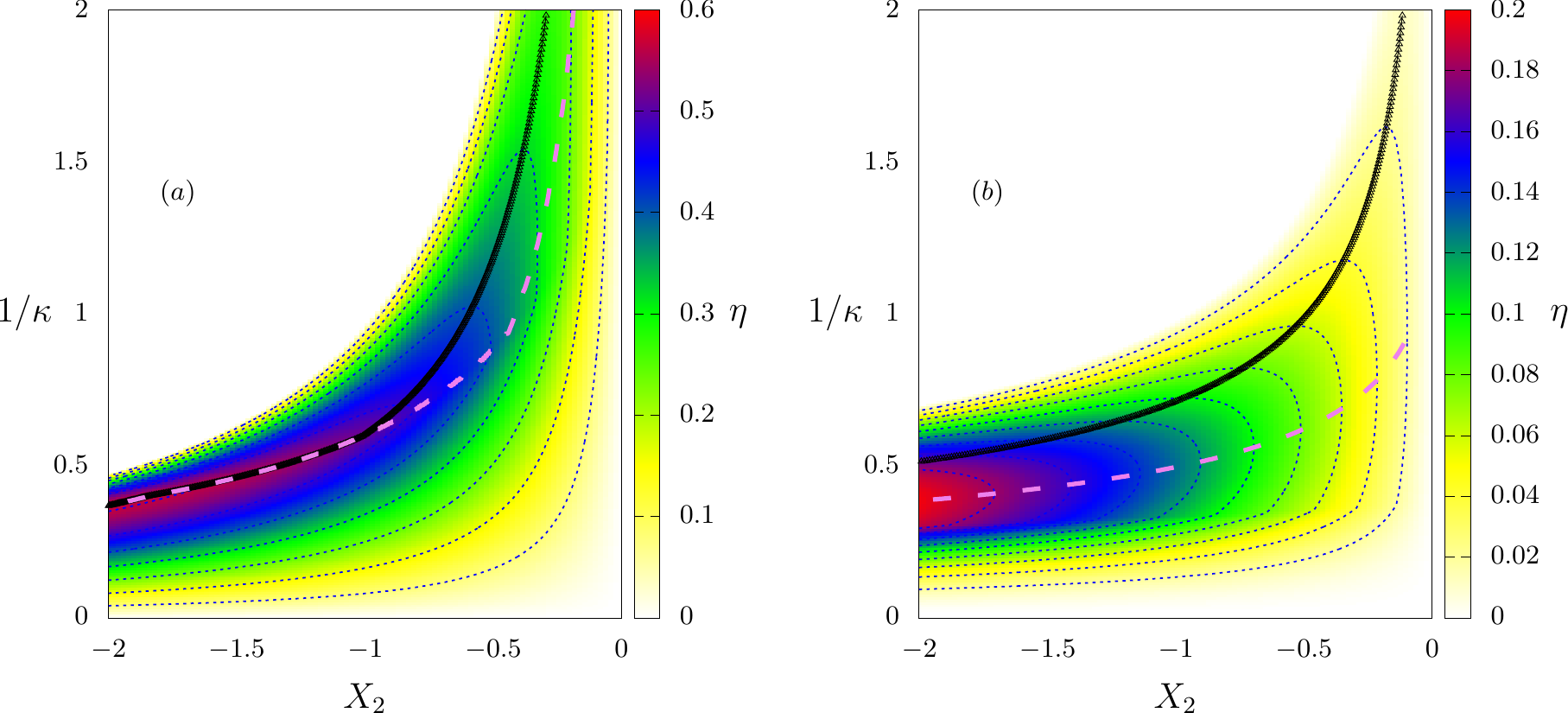}
      \caption{Depiction of efficiency as a function of 
        the inverse of the asymmetry coefficient $\kappa$ and
        the output force $X_{2}$, for constant $(a)$ and  linear $(b)$
        drivings, respectively. Solid and dashed lines denote the maximization with respect to $X_2$ and $\kappa$, respectively. Parameters:
        $\Gamma_1=2.0$ and $\Gamma_2=1.5$, $\tau=1, \; \gamma=1$ and $X_1=1$.  }
      \label{effe}
\end{figure}

Fig.~\ref{effe} presents heat maps of the efficiency for different temperature reservoirs as a function of the output force and asymmetry. By drawing a comparison with the isothermal work-to-work converter (Fig.~\ref{powere}), it reveals that the difference of temperature makes the choice of the optimization parameter (force strength or time asymmetry) more relevant. While both optimized lines lie almost on top of each other
for the isothermal case, Fig.~\ref{effe} shows that they are clearly distinct, particularly for the linear drivings. Another
point to be addressed concerns that high efficiencies  are restricted to larger $|X_2|$'s for constant drivings when temperatures are different. This contrasts to its extension to smaller values for isothermal reservoirs [the hot (red) region in Fig.~\ref{powere}$(b)$ is more spread than in Fig.~\ref{effe}$(a)$]. Conversely,  for linear drivings, the decrease of the efficiency extends for all values of $\kappa$ and $X_2$  when compared with the isothermal work-to-work converter [note that efficiency 
  in Fig.~\ref{powere}$(d)$ is 3 times larger than  Fig.~\ref{effe}$(b)$]. However, larger efficiencies in such
case is  obtained solely for larger values of $|X_2|$ 
under a certain range of $\kappa$. 

Lastly, we draw a comparison between 
the efficiency given by Eq. (\ref{eff1}) with Eq. (45) from Ref.~\cite{noa2020thermodynamics}, which is based on the ratio between the entropy production fluxes. Although
both expressions behave similarly and approach each other as $\Delta \Gamma\rightarrow 0$ (or $\Delta \Gamma<<1)$, it is worth mentioning that the latter
overestimates the efficiency as $\Delta \Gamma$  increases.

\section{Conclusions}

We introduced an alternative strategy for
optimizing the performance of  Brownian engines, based on the idea of asymmetric interaction time between the system (Brownian particle) and the thermal baths. Exact expressions
for thermodynamic quantities and their maximized values were obtained, irrespective the kind of
driving and asymmetry. The time asymmetry can always be tuned to obtain a gain larger than in the symmetric case. In addition to the improvement of the power output and efficiency, the time asymmetry also enlarges the range of forces for which the system operates as an engine. Another advantage of asymmetric times is that they can be conveniently chosen for compensating part of the  limitations due the machine design,
such as its operation period and the driving considered.

Results for constant and linear drivings confirm that the appropriate tuning of the asymmetry produce gains for the efficiency substantially larger than those achieved for the symmetric case. Contrariwise to usual machines, for which the heat flow due to the gradient of temperature is fundamental for the power extraction and enhancing the efficiency, in the present case the efficiency is higher for isothermal reservoirs. The reason for such  behavior concerns that the energy exchange between the Brownian particle and the different thermal reservoirs occurs in different stages. Since the heat transfer and the output force are uncoupled, the heat flux can not be converted into useful work. For instance,  one would require   drivings dependent of the velocity  in order to be able to extract work from heat in the present model. 
Although the robustness of our findings has been verified
for a few examples of drivings,  our approach  can be straightforwardly
extended for  other thermal machines, where in principle
similar findings are expected.  This is reinforced for recent results unveiling the importance of asymmetric times for optimizing the efficiency at maximum power of a quantum-dot thermal machine, which gain provides efficiencies larger than Curzon-Ahlborn \cite{PhysRevResearch.3.023194}.

We finish this paper highlighting a couple of perspectives. While in the present work we analyzed the maximization of the output power and efficiency with respect to the time asymmetry and the output force strength, keeping the other parameters of the machine fixed, it might be worth to study the maximization under different physical conditions, such as holding the dissipation or efficiency fixed. Finally, it might also be interesting to extend the role of asymmetric times for other kinds of drivings (e.g.\ velocity dependent drivings providing extraction of useful work from heat) as well as for massive Brownian particles (underdamped case) in order to compare their performances.
\section{Acknowledgment}
C. E. F acknowledges the financial support from FAPESP under grant
  2018/02405-1. AR thanks Pronex/Fapesq-PB/CNPq Grant No. 151/2018 and CNPq Grant No. 308344/2018-9.
\appendix
\section{Onsager coefficients and linear regimes}\label{app1}
In this appendix, we address the relation between  coefficients $\tilde{L}_{ij}$ and  Onsager coefficients $L_{ij}$. Our starting point is the steady state entropy production averaged over one period which is given by,
\begin{equation}
  \overline{\Pi} =  \frac{2 \gamma \kB}{m} \left(\frac{\overline{\dot{Q}}_1}{\Gamma_1} +  \frac{\overline{\dot{Q}}_2}{\Gamma_2} \right) = \overline{\Pi}_F + \overline{\Pi}_T.
  \label{pi}
\end{equation}
The coefficients $\tilde{L}_{ij}$ are straightforwardly obtained from $\overline{\Pi}_F$ performing the integration in Eq.~(\ref{epf}), which $\langle v_i\rangle(t)$'s are given by  Eq.~(\ref{eq:avvel}), as:
\begin{widetext}
\begin{equation}
\begin{split}
  {\tilde L}_{11} &= \frac{\gamma}{\tau }  \left[\frac{\left(e^{2 \gamma  (\tau-\tau_1)}-1\right) \hat{G}_1(\tau_1)^2}{\Gamma_2 \left(e^{\gamma  \tau}-1\right)^2}+\frac{\gamma}{\Gamma_1}  \int_0^{\tau_1} \frac{2 e^{-2 \gamma  t} \left[\left(e^{\gamma  \tau }-1\right) \hat{G}_1(t)+\hat{G}_1(\tau_1)\right]^2}{\left(e^{\gamma  \tau }-1\right)^2} \, dt\right] \\
  {\tilde L}_{22} &= \frac{\gamma}{\tau }  \left[\frac{\left(1-e^{-2 \gamma  \tau_1}\right) \hat{G}_2(\tau)^2}{\Gamma_1 \left(e^{\gamma  \tau }-1\right)^2} + \frac{\gamma}{\Gamma_2}  \int_{\tau_1}^{\tau } \frac{2 e^{-2 \gamma  t} \left[\left(e^{\gamma  \tau }-1\right) \hat{G}_2(t)+\hat{G}_2(\tau)\right]^2}{\left(e^{\gamma  \tau }-1\right)^2} \, dt\right], \\
  {\tilde L}_{12}+{\tilde L}_{21} &= \frac{2 \gamma  e^{-\gamma  \tau_1} \hat{G}_1(\tau_1) \hat{G}_2(\tau) }{\tau  \left(e^{\gamma  \tau}-1\right)^2}\left[ \frac{\sinh (\gamma  \tau_1)}{\Gamma_1}  +  \frac{\sinh (\gamma  (\tau -\tau_1))}{\Gamma_2}\right] \\
  &+\frac{2 \gamma ^2}{\Gamma_1 \Gamma_2\tau \left (e^{\gamma  \tau }-1\right)} \left[ \Gamma_2 \hat{G}_2(\tau) \int_0^{\tau_1} \hat{G}_1(t) e^{-2 \gamma  t} \, dt + \Gamma_1 \hat{G}_1(\tau_1) \int_{\tau_1}^{\tau } \hat{G}_2(t) e^{\gamma  (\tau -2 t)} \, dt\right],
\end{split}
\label{l11}
\end{equation}
\end{widetext}
where $\hat{G}_i(t) = \int_{\tau_{i-1}}^t g_i(t^\prime) dt^\prime$. 
For equal temperatures $\Gamma_1 = \Gamma_2 = \Gamma$, $\overline{\Pi}$ reduces to the following expression:
\begin{equation}
  \begin{split}
  \overline{\Pi} &= \overline{\Pi}_F =-\frac{2 \gamma \kB}{m \Gamma}  \left(\overline{\dot{W}}_1 + \overline{\dot{W}}_2\right)\\
  & = L_{11} X_1^2 + \left(L_{12} + L_{21} \right) X_1 X_2 + L_{22} X_2^2.
  \end{split}
\end{equation}
Hence,  for isothermal reservoirs the entropy production can be written in terms of the Onsager coefficients even in the non-linear (force) regime and
thereby  $\tilde{L}_{ij} = L_{ij}$.
Conversely, for the thermal linear regime, it is convenient to express $\Gamma_1$ and $\Gamma_2$
in terms of the difference of temperatures  $\Gamma_1 = \Gamma - \Delta \Gamma$ and $\Gamma_2 = \Gamma + \Delta \Gamma$. In such case, Eq. (\ref{pi}) becomes
\begin{equation}
   \overline{\Pi}\approx \frac{2 \gamma \kB}{m \Gamma} \left[ -\left(\overline{\dot{W}}_1 + \overline{\dot{W}}_2\right) +  \left (\overline{\dot{Q}}_1 - \overline{\dot{Q}}_2\right) \frac{\Delta \Gamma}{\Gamma}\right].
   \label{eq:phi}
\end{equation}
Let us assume that ${\tilde L}_{ij}$ can be expanded
in power series of the temperature difference, ${\tilde L}_{ij} = L_{ij}^{(0)} + L_{ij}^{(c)} \Delta \Gamma$,
where $L_{ij}^{(0)}$ is the coefficient for $\Gamma_1 = \Gamma_2 = \Gamma$ and $L_{ij}^{(c)}$ is the first order correction. In terms of such coefficients, the average entropy production $\overline{\Pi}$ is given by
\begin{equation}
  \begin{split}
    \label{eq:pi}
   \overline{\Pi} &= \overline{\Pi}_F + \overline{\Pi}_T\\
   &=\left[ L_{11}^{(0)} X_1^2 + \left(L_{12}^{(0)}+L_{21}^{(0)} \right) X_1 X_2 + L_{22}^{(0)} X_2^2 \right]+\\
   &+ \left[ L_{11}^{(c)} X_1^2 + \left(L_{12}^{(c)}+L_{21}^{(c)} \right) X_1 X_2 + L_{22}^{(c)} X_2^2 \right] \Delta \Gamma\\& + \frac{4 L_{\Gamma \Gamma}}{\Gamma^2} \left(\Delta \Gamma \right)^2.
  \end{split}
\end{equation}
By comparing Eqs.~(\ref{eq:phi}) and (\ref{eq:pi}), it follows that
    \begin{equation}
      L_{11}^{(0)} X_1^2 + \left(L_{12}^{(0)}+L_{21}^{(0)} \right) X_1 X_2 + L_{22}^{(0)} X_2^2  = -\frac{2 \gamma \kB}{m \Gamma} \left(\overline{\dot{W}}_1 + \overline{\dot{W}}_2\right),
    \end{equation}
    and hence Onsager coefficients $L_{ij}$'s correspond to 0-th order
    coefficients $L_{ij}^{(0)}$'s
    evaluated  from $\overline{\Pi}_F$. Once again, they
 do not depend on $\Delta \Gamma$, since $\overline{\dot{W}}_i$ does not depend on the temperature at all.
 
    In the true linear regime (both  temperature gradient and force strength are small), the correction of $\overline{\Pi}_F$ is of third order ($X_i X_j \Delta \Gamma$), thus it can be neglected. Hence, the entropy production components $\overline{\Pi}_F$ and $\overline{\Pi}_T$ are approximately 
    \begin{equation}
      \overline{\Pi}_F \approx  -\frac{2 \gamma\kB}{m \Gamma} \left(\overline{\dot{W}}_1 + \overline{\dot{W}}_2\right), 
      \label{pif}
    \end{equation}
    and
     \begin{equation}
       \overline{\Pi}_T \approx \frac{4 L_{\Gamma \Gamma}}{\Gamma^2} \left(\Delta \Gamma \right)^2,
        \label{pit2}
    \end{equation}
respectively. In addition, the coefficients $\tilde{L}_{ij}$ and $L_{ij}$ are approximately equal $\tilde{L}_{ij} \approx L_{ij}$.

\section{Constant drivings}
\label{sec:constforce}

For the machine operating at constant drivings, defined by the forces from Eqs.~(\ref{eq:constforce1}) and~(\ref{eq:constforce2}), the  velocities $\langle v_i\rangle(t)$'s are given by
\begin{eqnarray}\label{VMC}
  \langle v_{1}\rangle(t)&=& \frac{X_1}{\gamma }+\frac{e^{-\gamma  (t-\tau_1)}-e^{-\gamma  (t-\tau)})}{e^{\gamma  \tau }-1}\frac{X_1-X_2}{\gamma},\\
  \langle v_{2}\rangle(t)&=& \frac{X_2}{\gamma }+\frac{e^{-\gamma  (t-\tau -\tau_1)}-e^{-\gamma (t-\tau)}}{e^{\gamma  \tau }-1}\frac{X_1-X_2}{\gamma},
\end{eqnarray}
for $i=1$ and $2$, respectively.
The associated Onsager coefficients are straightforwardly obtained from Eq.~(\ref{l21ew}) and are given by
\begin{eqnarray}
  L_{11} &=&  \frac{2 \tau_1}{\Gamma_1 \tau} - L_{12}, \nonumber \\
  L_{22} &=&  \frac{2 \tau_2}{\Gamma_2 \tau} - L_{21}, \nonumber \\
  L_{12} &=& \frac{4 \csch \left(\frac{\gamma  \tau }{2}\right) \sinh \left(\frac{\gamma \tau_1}{2}\right) \sinh \left(\frac{1}{2} \gamma  \tau_2\right)}{\gamma  \Gamma_1 \tau },\\
  L_{21} &=& \frac{4 \csch \left(\frac{\gamma  \tau }{2}\right) \sinh \left(\frac{\gamma \tau_1}{2}\right) \sinh \left(\frac{1}{2} \gamma  \tau_2\right)}{\gamma  \Gamma_2 \tau}.\nonumber 
\end{eqnarray}
Furthermore, for isothermal reservoirs, $L_{12}$ and $L_{21}$ are equal for any value of asymmetry parameter $\kappa=\tau_1/\tau_2$.

 \section{Linear drivings}
\label{sec:linforce}
Similarly to the constant drivings model, the average velocities for the linear driving model [defined by Eqs.~(\ref{eq:linforce1}) and~(\ref{eq:linforce2})] is obtained from Eq.~(\ref{eq:avvel}) and are given by
\begin{widetext}
\begin{equation}
\begin{split}
     \langle v_{1} \rangle(t)  =\frac{1}{\gamma}\left\{X_{1}(\gamma t-1)+\frac{e^{-\gamma t}}{e^{\gamma\tau}-1}\left\{X_{1}\left[e^{\gamma\tau}+e^{\frac{\gamma\kappa\tau}{1+\kappa}}\left(\frac{\gamma\kappa\tau}{1+\kappa}-1\right)\right]-              X_{2}[e^{\frac{\gamma\kappa\tau}{1+\kappa}}+e^{\gamma\tau}\left(\frac{\gamma\tau}{1+\kappa}-1\right)]\right\}\right\},
\end{split}
\end{equation}
and
\begin{equation}
\begin{split}
     \langle v_{2} \rangle(t)  =\frac{1}{\gamma}\left\{X_{2}[1-\gamma\left(t-\frac{\kappa\tau}{1+\kappa}\right)]+\frac{e^{-\gamma (t-\frac{\kappa\tau}{1+\kappa})}}{e^{\gamma\tau}-1}\left\{X_{1}[e^{\frac{\gamma\tau}{1+\kappa}}+e^{\gamma\tau}\left(\frac{\gamma\kappa\tau}{1+\kappa}-1\right)]-
                               X_{2}[e^{\frac{\gamma\tau}{1+\kappa}}\left(\frac{\gamma\tau}{1+\kappa}-1\right)+e^{\gamma\tau}]\right\}\right\}.
\end{split}
\end{equation}
Likewise,  Onsager coefficients $L_{ij}$'s are also straightforwardly calculated from Eq.~(\ref{l21ew}) and read
\begin{equation}
  \begin{split}
  L_{11}&=\frac{2 \gamma ^3 \tau_1^3+\left[6-3 \gamma ^2 \tau_1^2\right] \coth \left(\frac{\gamma  \tau }{2}\right)+6 \csch\left(\frac{\gamma  \tau }{2}\right) \left[\gamma  \tau_1 \sinh \left(\frac{\gamma  (\tau_1-\tau_2)}{2}\right)-\cosh \left(\frac{\gamma  (\tau_1-\tau_2)}{2}\right)\right]}{3 \gamma  \Gamma  \tau }.\\
L_{22}&= \frac{2 \gamma ^3 \tau_2^3 +\left[6 - 3\gamma ^2 \tau_2^2-6 \cosh (\gamma  \tau_1)\right]\coth \left(\frac{\gamma  \tau }{2}\right) +6 \gamma \tau_2 \csch\left(\frac{\gamma  \tau }{2}\right) \sinh \left(\frac{\gamma  (\tau_2-\tau_1)}{2}\right)+6 \sinh (\gamma  \tau_1)}{3 \gamma  \Gamma  \tau },\\
L_{12}&=\frac{2}{\gamma\tau\Gamma_1(1-e^{\gamma\tau})}\left[1+\gamma\tau_1-e^{\gamma\tau_1}\right]\left[1-e^{\gamma\tau_2}\left(1-\gamma\tau_2\right)\right],\\
  L_{21}&=\frac{2}{\gamma\tau\Gamma_2(1-e^{\gamma\tau})}\left[1+\gamma\tau_2-e^{\gamma\tau_2}\right]\left[1-e^{\gamma\tau_1}\left(1-\gamma\tau_1\right)\right].
  \end{split}
\end{equation}
Notably, contrasting to the constant drivings case, coefficients $L_{12}$ and $L_{21}$ are different from each other when $\Gamma_1=\Gamma_2$. Only
for  symmetric switching times ($\tau_1 =\tau_2$), it turns out that $L_{12}=L_{21}$.
\end{widetext}
\bibliographystyle{apsrev4-1}
\bibliography{references}
 
\end{document}